\documentclass{iopart}

\usepackage{iopams}
\usepackage{graphicx}
\usepackage{setstack}

\newcommand{\eqref}[1]{(\ref{#1})}

\begin{document}

\title{Hartree shift in unitary Fermi gases}
\author{J. J. Kinnunen}
\address{Department of Applied Physics, Aalto University School of Science and Technology, Finland}
\ead{jjkinnun@gmail.com}
\begin{abstract}
The Hartree energy shift is calculated for a unitary Fermi gas. By including the momentum dependence of 
the scattering amplitude explicitly, the Hartree energy shift remains finite even at unitarity. 
Extending the theory also for spin-imbalanced systems allows
calculation of polaron properties. The results are in good agreement with more 
involved theories and experiments.
\end{abstract}


\maketitle

\section{Introduction}

Strongly interacting atomic gases have proved to be an excellent platform 
for studying many-body quantum phenomena~\cite{Bloch2008a}. The purity and the
diluteness of the gases allow the description of the system using well
understood microscopic models~\cite{Holland2001a,Ohashi2002a,Falco2005a} and a host of 
theoretical tools have been applied for solving stationary and dynamic
properties of the system~\cite{Georges1996a,Carlson2003a,Vidal2003a,Bulgac2009a}.

Despite the multitude of more exact and involved theoretical tools, mean-field
theories form an important class of theories due to their inherent simplicity and
the applicability to three-dimensional systems~\cite{Chen2005a}. Surprisingly, 
these mean-field theories have proved to be able to describe qualitatively even strongly interacting
systems~\cite{Giorgini2008a} although finite temperatures and polarized gases have
proved challenging. Indeed, critical temperatures and critical polarizations for the
superfluid-normal transitions are exaggerated in a mean-field theory. These problems
are caused by the difficulty of describing the non-superfluid state of the gas
properly, the state which is often referred to as the 'pseudogap phase'~\cite{Chen2005a}. 
Indeed, the non-superfluid state in the BCS mean-field theory is a non-interacting finite
temperature Fermi sphere in which all interaction effects are absent; a strong
assumption for a strongly interacting system.

The natural question is whether the simple BCS mean-field theory can be improved
to include interaction effects in the normal state at least approximately. Obviously,
there are a lot of extensions of the simple BCS theory~\cite{Nozieres1985a,Engelbrecht1992a,deMelo1993a,Haussmann1994a,Perali2002a,Stajic2004a,
Chen2005a,Haussmann2009a,Tsuchiya2009a,Hu2010a,Watanabe2010a,Gubbels2011a} that incorporate
fluctuations in the BCS model and can describe the non-superfluid state properly.
Most of the methods are however numerically very heavy and the simplicity of the
BCS mean-field theory has been lost. The approach in this work is quite different.
Instead of considering fluctuations around the BCS theory, I neglect all pairing
effects altogether and concentrate on one key approximation done in the BCS theory
when in the strongly interacting regime, namely the omission of the Hartree energy
shift. The idea is not new as pairing correlations are often neglected when
considering systems at high temperatures but here I will relax approximations
often done to the Hartree energy shift~\cite{Chiacchiera2009a}. The result is a self-consistent theory which
is simple but sufficiently powerful to describe a wide range of phenomena. 

The structure of this work is the following: in Section~\ref{sec:hartree_initial} I will first
provide a definition for the Hartree energy shift for a strongly interacting gas. 
In Section~\ref{sec:scattering} the main points of the standard 
scattering calculation for deriving the many-body scattering T-matrix are reviewed. The details of the 
derivation can be found in~\ref{app:scatamp}. 
In Section~\ref{sec:hartree} the Hartree self-energy will be calculated for a balanced 
two-component Fermi gas at unitarity. In Section~\ref{sec:hartreeforpolarized} the theory
is generalized for spin-imbalanced systems and the Hartree self-energy and the effective mass are calculated for a polaron. 
In Section~\ref{sec:contact} the theory will be further extended for calculating the occupation numbers at the high momentum limit 
and the Tan's contact coefficient will be solved both for a balanced system and for a polaron. The key approximations
are reviewed and the implications of the theory are discussed in Section~\ref{sec:discussion}.
Finally, the key findings are summarized in Section~\ref{sec:summary}.

\section{The Hartree shift}
\label{sec:hartree_initial}

The effect of atom-atom interactions enter the single-particle Green's function $G(P)$
through the self-energy $\Sigma(P)$ according to the Dyson equation
\begin{equation}
   G(P) = G_0(P) + G_0(P) \Sigma(P) G(P) = \frac{1}{G_0(P)^{-1} - \Sigma(P)},
\end{equation}
where the non-interacting Green's function 
$G_0(P) = G_0({\bf p},p_0) = 1/(p_0 - \epsilon_p)$. 
The standard approximation for the self-energy used in the context
of dilute atomic gases is the Hartree-Fock self-energy written using
a scattering T-matrix
\begin{equation}
   \Sigma(P) = -i \int  \frac{dK}{\left(2\pi\right)^4}\, G(K) \Gamma (K,P;P,K).
\label{eq:hfselfenergy}
\end{equation}
This describes the self-energy of a particle with four-momentum $P$, 
interacting with a particle of four-momentum $K$ drawn from the bath. The
interaction is described by the \emph{many-body scattering T-matrix} 
$\Gamma (K,P;P,K)$ which promptly returns the particles back to the initial
state, thus returning the particle with the four-momentum $K$ to the bath and
the initial particle to the state $P$. The self-energy utilises only the diagonal
part of $\Gamma$, but the T-matrix will eventually be needed for a more general 
case as well. The above form assumes that the two interacting particles are distinguishable and thus it
suffices for describing interspecies interactions in two-component Fermi gases. For indistinguishable 
particles one should add also the exchange interaction channel.

Replacing the many-body scattering T-matrix by a constant 
$\Gamma \approx \frac{4\pi\hbar^2}{m} a$, where $a$ is the s-wave scattering
length, yields the well known result for the Hartree energy shift 
in the weakly interacting limit
\begin{equation}
  \Sigma_\mathrm{Hartree} = \frac{4\pi\hbar^2}{m} a n,
\label{eq:weakhartree}
\end{equation}
where $n$ is the atom density. However, this formula does not work
in the strongly interacting, or unitary, limit where $a \rightarrow \infty$. Indeed,
at unitarity the only relevant length scale
in a dilute Fermi gas is the Fermi momentum $k_\mathrm{F}$ and the
Hartree energy shift must become independent of $a$~\cite{Heiselberg2001a,Giorgini2008a}. Simply
replacing $a$ by $1/k_\mathrm{F}$ would yield as the Hartree energy
shift $4/3\pi E_\mathrm{F} \approx 0.42\,E_\mathrm{F}$, which is already
in a rather good agreement with the values obtained for the Hartree-like
energy shift in a Monte Carlo study~\cite{Magierski2009a} and a self-consistent
Green's function theory~\cite{Haussmann2009a}.

However, such a naive argument raises more questions than answers, and
the purpose of this work is to make the argument rigorous and especially
clarify the physics behind it. However, the main problem with discussing
the Hartree shift in the strongly interacting regime or at unitarity is
that there is no good definition for it. While in the weakly interacting
limit the Hartree shift can be considered as an energy shift linear in
density $n$, at unitarity it is proportional to $n^{2/3}$. Another important
issue to notice is that the Hartree shift is not constant, i.e. equal
for all atoms, even in the weakly interacting limit. This follows from 
the breaking of the translational symmetry due to the presence of the
Fermi sphere~\cite{Pekker2011a} coupled with the momentum dependence of the two-particle 
scattering amplitude, which essentially says that the interaction strength 
between two atoms becomes weaker when the relative momentum increases. 
For weakly interacting atoms this momentum
dependence is not relevant as long as $|k_\mathrm{F} a| \ll 1$, but 
in strongly interacting systems one can hardly expect the Hartree shift
to be a constant. 

Since the Hartree shift in the strongly interacting systems has so few 
properties in common with the weakly interacting limit, one can ask whether
it even makes sense to talk about Hartree shifts at unitarity. Indeed,
many advanced methods designed for describing strongly interacting
systems do not separate the Hartree contribution at all~\cite{Haussmann2009a,Perali2002a}
or identify it only as a part of the total grand potential~\cite{Diener2008a}, which
is a global quantity and not a quantity directly relevant for single-particle excitations (and hence
for the single-particle Green's function $G(K)$).
However, in simpler mean-field theories, such as the BCS theory applied to the
strongly interacting regime~\cite{Eagles1969,Leggett}, the Hartree shift is explicitly
\emph{neglected} and consequently also easily identified~\cite{Giorgini2008a}. The approach of
this work is to omit the terms traditionally \emph{included} in the BCS theory
and instead consider the Hartree shift in more detail. In the future, it will be
interesting to consider approximative methods that combine the two, the BCS 
theory and the Hartree shift.
 
Solving the $k_0$ frequency integral in Eq.~\eqref{eq:hfselfenergy} (by closing the integration path in a 
semi-circle around the upper half of the complex plane) yields contributions 
from poles in the Green's function $G$ but also from possible poles in $\Gamma$ that are
located in the upper half of the complex plane. The self-energy can thus be separated as
\begin{equation}
  \Sigma (P) = \Sigma_\mathrm{Hartree} + \Sigma_\mathrm{pairs},
\label{eq:selfenergyseparation}
\end{equation}
where the Hartree part $\Sigma_\mathrm{Hartree}$ of the self-energy 
corresponds to the part given by
the pole in the Green's function and the pair part $\Sigma_\mathrm{pairs}$ is 
given by the poles in the T-matrix. This is the definition
of the Hartree self-energy used in this work: it is the part of the self-energy
arising from the pole in the Green's function in Eq.~\eqref{eq:hfselfenergy}. 
In addition, by the Hartree \emph{shift}
I will refer to the real part of the Hartree self-energy $\mathrm{Re}\,\Sigma_\mathrm{Hartree}$. 
Mathematically speaking this separation to Hartree and pair self-energies is 
very easy to do as the contributions from the poles add up exactly
in such a way when doing contour integration. Physically speaking the poles in the Green's 
function give the occupation
numbers and, eventually, the atom density. The poles of the T-matrix correspond
to many-body bound states or molecular states if the two-body
interaction potential supports such states. The simplest theory for describing 
these many-body bound states would be the BCS theory, in which case the Cooper 
pair formation is signaled by the appearance of a pole in the many-body T-matrix 
at zero center-of-mass momentum and energy. However, the pole of the Green's function 
is neglected in the BCS self-energy for strong interactions, as the dominant term is 
considered to be the pole in the T-matrix describing the Cooper pair formation. 
The above definition for the Hartree self-energy is very similar to the self-energy used in 
Ref.~\cite{Chiacchiera2009a}, and notice also that it yields 
Eq.~\eqref{eq:weakhartree} in the weakly interacting limit when the T-matrix is
replaced by the constant $4\pi\hbar^2 a/m$.

\section{Many-body scattering T-matrix at unitarity}
\label{sec:scattering}

I will start by reviewing the derivation of the many-body scattering 
T-matrix. While the final result is already well known and widely used, 
the actual derivation is not often shown, with the calculation often 
starting from the two-body scattering T-matrix instead of the actual 
atom-atom interaction potential. However, going through the actual 
derivation may help to understand the physics of the Hartree self-energy and 
where the various features in the many-body scattering T-matrix arise from. 
The detailed calculation can be found in~\ref{app:scatamp}, but here 
in the main text I will do the key approximations and show only the 
main results.
 
\subsection{Inelastic scattering amplitude}

The first key result from the scattering calculation in~\ref{app:scatamp}
is the elastic scattering amplitude for a short range potential
\begin{equation}
  f({\bf s},{\bf k}) = f(k) \approx \frac{-a}{1+ika},\quad \mathrm{for} \, k = s
\end{equation}
which describes the scattering of two particles with the incoming and outgoing
(relative) momenta ${\bf k}$ and ${\bf s}$, respectively. For short
range potentials, the scattering length $a$ is enough to parametrize
the two-body scatterings in vacuum. As described in~\ref{app:scatamp}, I will
assume that the \emph{inelastic} scattering amplitude has the same form, i.e.
$f({\bf s,k}) = f(k)$ for all ${\bf s},{\bf k}$. While such assumption
is not true in general, it is consistent with the generalized
optical theorem for inelastic scatterings in Eq.~\eqref{eq:opticaltheory}
at unitarity $a \rightarrow \infty$.

At low energies, $k \approx 0$, the scattering amplitude is constant $-a$, but 
the momentum dependence and the associated decay of the real part of 
the scattering amplitude,
\begin{equation}
    \mathrm{Re}\, f(k) = \frac{-a}{1+(ka)^2},
\end{equation}
implies the well known fact that the interaction strength decreases when
the relative momentum of the two particles increases. For interaction potentials
with a finite range, the decay is even faster. This means that a very
fast atom does not feel interaction effects and all kinds of pair formation
and Hartree energy shifts are suppressed. The idea of a constant Hartree
shift thus holds only for a limited range of momenta, with the characteristic
scale given by the inverse of the scattering length $1/a$. Obviously,
in strongly interacting gases the condition $ka \ll 1$ is easily breached.

The momentum dependence of the Hartree energy shift has been observed in 
Bragg spectra of strongly interacting Bose-Einstein condensates~\cite{Papp2008a,Kinnunen2009a,Ronen2009a}.
However, compared to strongly interacting fermionic gases, the associated 
momenta $k$ and scattering lengths $a$ in that experiment were still rather meager.
In this work I consider only the two limits: the weakly interacting 
limit $f({\bf s},{\bf k}) \approx -a$ and the unitarity $f({\bf s},{\bf k}) = i/k$,
but the theory is quite easily generalized to intermediate interactions as well.

\subsection{Two-body scattering T-matrix}

The scattering amplitude describes stationary scattering states but in
order to calculate the self-energy the two-body scattering
T-matrix in Eq.~\eqref{eq:gamma0vsscatamp} is needed
\begin{eqnarray}
\fl   \Gamma_0({\bf s}, {\bf k}; P) =& -\frac{4\pi \hbar^2}{m} f({\bf s}, {\bf k}) \\
\fl &+\left( \frac{4\pi \hbar^2}{m} \right)^2 \int \frac{d{\bf q}}{(2\pi)^3} f({\bf s}, {\bf q}) \left( \frac{1}{2p_0-2\epsilon_p - 2\epsilon_q} + \frac{1}{2\epsilon_q - 2\epsilon_{k} - i\eta} \right) f ({\bf k}, {\bf q})^*.\nonumber
\label{eq:gamma0vsscatamp2}
\end{eqnarray}
The physical interpretation of the two-body scattering T-matrix is very
simple. The scattering amplitude $f({\bf s},{\bf k})$ obtained from the standard scattering
calculation describes stationary solutions of the scattering problem,
and thus the energies of the incoming and outgoing particles are equal
(there is only one 'energy' for each eigenstate of the Hamiltonian). 
This implies that the energy is not a free parameter in the scattering
calculation. The two-body scattering T-matrix $\Gamma_0$ now generalizes this scattering
amplitude into the case in which the energy is a free parameter. This is important
when considering for example scattering processes where the particles can
exchange energy with the environment (say, a probing radio-frequency or laser
field) but also when the energy is not a well defined concept such as 
in the case of virtual intermediate states.

In the weakly interacting limit the two-body scattering T-matrix simplifies into
\begin{eqnarray}
  \Gamma_0({\bf s}, {\bf k}; P) =& -\frac{4\pi \hbar^2a}{m} + \mathcal{O}\left( a^2 \right).
\end{eqnarray}
At unitarity it becomes
\begin{equation}
\fl   \Gamma_0({\bf s}, {\bf k}; P) 
= -\frac{4\pi \hbar^2}{m} \frac{i}{k} +
\left( \frac{4\pi \hbar^2}{m} \right)^2 \int \frac{d{\bf q}}{(2\pi)^3} \frac{1}{q^2} \left( \frac{1}{2p_0 - 2\epsilon_p - 2\epsilon_{q}} + \frac{1}{2\epsilon_{q} - 2\epsilon_{k} - i\eta} \right).
\end{equation}
The first important point to notice is that the two-body scattering T-matrix does not depend on the momentum ${\bf s}$ at all,
i.e. the approximation done for the scattering amplitude, $f({\bf s},{\bf k}) \approx f(k)$, 
carries on to the two-body scattering T-matrix.

The integrand is spherically symmetric and the integral is easily evaluated
\begin{eqnarray}
\fl   \Gamma_0({\bf s}, {\bf k}; P) &= -\frac{4\pi \hbar^2}{m} \frac{i}{k} +
\left( \frac{4\pi \hbar^2}{m} \right)^2 \frac{4\pi}{(2\pi)^3} \int dq \left( \frac{1}{2p_0 - 2\epsilon_p - 2\epsilon_{q}} + \frac{1}{2\epsilon_{q} - 2\epsilon_{k} - i\eta} \right) \nonumber \\
\fl &= -\frac{4\pi \hbar^2}{m} \frac{i}{k} + \left( \frac{4\pi \hbar^2}{m} \right)^2 \frac{4\pi}{(2\pi)^3} \int dq \, \frac{1}{2p_0 - 2\epsilon_p - 2\epsilon_{q}} + i \frac{4\pi \hbar^2}{m} \frac{1}{k}  \nonumber \\
\fl &= 4 \frac{\hbar^4}{m^2} \int dq \, \frac{1}{\epsilon+i\eta - \epsilon_p - \epsilon_{q}} = \frac{4\pi\hbar^2}{m} \sqrt{\frac{\hbar^2}{2m \left(\epsilon_p - \epsilon \right)}},
\label{eq:gamma0unit}
\end{eqnarray}
where in the last line I have done the analytic continuation $p_0 = \epsilon + i\eta$ with 
an infinitesimally small imaginary part $\eta \rightarrow 0$. The two-body scattering 
T-matrix is thus either real or pure imaginary, depending on the sign of $\epsilon_p-\epsilon$.
Notice also that the two-body scattering T-matrix does not depend on either momenta
$s$ or $k$ but only on the center-of-mass momentum ${\bf p}$ and energy $p_0$. 
However, one should notice that this property is somewhat illusionary: for
an elastic scattering the two-body scattering T-matrix is identical to the scattering amplitude $f(k)$,
which depends on the relative momentum ${\bf k}$ \emph{instead} of the
center of mass momentum ${\bf p}$. Indeed, for an elastic scattering
$\epsilon-\epsilon_p = \epsilon_k$ and the two-body scattering T-matrix regains the
form used for the scattering amplitude.

\subsection{Many-body scattering T-matrix}

The many-body scattering T-matrix can be expressed using the two-body
scattering T-matrix as in Eq.~\eqref{eq:gamma})
\begin{eqnarray}
   \Gamma ({\bf s},{\bf k};P)= &\Gamma_0 ({\bf s},{\bf k};P) + \int \frac{d{\bf q}}{(2\pi)^3} \, \Gamma_0 ({\bf s},{\bf q};P) \tilde \chi ({\bf q};P) \Gamma ({\bf q}, {\bf k};P).
\end{eqnarray}
Using the knowledge that the two-body scattering T-matrix depends only on the center-of-mass four-momentum $P$ allows writing this as
\begin{eqnarray}
   \Gamma (P)= \frac{\Gamma_0(P)}{1- \Gamma_0 (P) \int \frac{d{\bf q}}{(2\pi)^3} \,\tilde \chi ({\bf q};P)},
\end{eqnarray}
Combining this with the result obtained for $\Gamma_0$ in Eq.~\eqref{eq:gamma0unit} yields
\begin{equation}
  \Gamma (P) = -\frac{1}{g(P)}
\label{eq:gammavsg}
\end{equation}
where 
\begin{equation}
\fl   g(P) = i\int \frac{dQ}{(2\pi)^4} G(P+Q) G(P-Q) + \int \frac{d{\bf q}}{(2\pi)^3} \frac{\mathcal{P}}{2p_0 - 2\epsilon_q - 2\epsilon_p},
\label{eq:gunitarity}
\end{equation}
where $\mathcal{P}$ signifies the principal value integration. 
This is exactly the same as the standard T-matrix at unitarity $a \rightarrow \infty$~\cite{Chen2005a},
and as pointed out at the end of~\ref{app:scatamp}, the many-body scattering T-matrix yields
the standard result whenever the two-body scattering T-matrix does not explicitly depend 
on the relative momenta ${\bf s}$,${\bf k}$. 


In order to proceed further, I will make an ansatz for the Green's function $G(P)$. The choice
of the ansatz is probably the most important approximation in this work and hence it must
be chosen with care. I assume that the Green's functions have the form
\begin{equation}
  G(P) = \frac{n(\epsilon_p)}{p_0^* - \epsilon_{\bf p} - h^*(p)} + \frac{1-n(\epsilon_p)}{p_0 - \epsilon_{\bf p} - h(p)},
\label{eq:green}
\end{equation}
where $n(\epsilon_p)$ is the occupation probability for a single-particle state with energy $\epsilon_p$ (or momentum $p$). 
The first term, the advanced Green's function, describes a hole excitation with energy (dispersion) $\epsilon_p + h^*(p)$ and 
the second term, the retarded Green's function, describes a particle excitation. The complex conjugation of $h(p)$ in the hole 
branch guarantees that the poles corresponding to holes remain in the upper half of the complex plane and the particle poles 
in the lower half (the imaginary part of $h(p)$ must thus be negative). The key approximations in this ansatz are that the 
self-energies $h(p)$ for the advanced and retarded parts are the same and that they do not depend on the frequency $p_0$. 
These approximations exclude a lot of physics from entering the formalism, perhaps the most important piece missing being 
the formation of an excitation gap. Indeed, the
BCS Green's function would require a self-energy of the form $\Sigma(P) = -G_0(-P)\Delta^2$, where $\Delta$ is the
BCS excitation gap and $G_0(P)$ is the non-interacting Green's function, which clearly depends on the frequency $p_0$. 
However, the approximations fit with the picture of a momentum-dependent Hartree shift, with each Green's function pole
being simply shifted by the amount $\mathrm{Re}\, h(p)$. By allowing complex values of $h(p)$ the formalism also incorporates
lifetime effects, broadening the spectral function. The ansatz in Eq.~\eqref{eq:green} is thus sufficient
for describing Hartree-like effects. Taking into account also the pairing effects would generalize the present
theory in the direction of the pair-fluctuation theory in Ref.~\cite{Perali2002a}. However, here I will concentrate
on the Hartree effect and ignore the pairing contribution.

The ideal zero-temperature Green's function can be obtained from the ansatz by choosing $n(x) = 1-\theta(x)$ and
non-interacting finite temperature Green's function by using the Fermi-Dirac distribution. Below I will also consider 
occupation numbers of the form given by the Bogoliubov coefficients 
\begin{equation}
  n(\epsilon_p) = v_p^2 = \frac{1}{2} \left( 1 - \frac{\epsilon_p-\mu}{\sqrt{\left(\epsilon_p - \mu\right)^2 + \Delta^2}}\right)
\label{eq:bpcansatz}
\end{equation}
in order to have an ansatz that satisfies the Bethe-Peierls condition for the two-body physics~\cite{Giorgini2008a, Cazalilla2010a}.
However, despite the similarity with the BCS Green's function, this does not result in the emergence 
of an excitation gap. Instead, the parameter $\Delta$ is simply a measure of the broadening of the Fermi surface.
In order to emphasize this difference with the true BCS Green's function, but also to show the purpose of the
ansatz, I will below refer to the ansatz in Eq.~\eqref{eq:bpcansatz} by the 'BPC ansatz'. The BPC ansatz will
be discussed further in Section~\ref{sec:contact}.
As will be seen later, the Hartree energy shifts are not very sensitive to the exact form of the occupation
number ansatz. 

The regularized pair susceptibility can now be evaluated using contour integrals yielding
\begin{equation}
\fl   g(P) = \int \frac{d{\bf q}}{(2\pi)^3} \left[ \frac{n(-\epsilon_+)n(-\epsilon_-)}{2p_0 - \varepsilon_+ - \varepsilon_-} - \frac{n(\epsilon_+) n(\epsilon_-)}{2p_0^* - \varepsilon_+^* - \varepsilon_-^*} - \frac{\mathcal{P}}{2p_0 - \epsilon_+ - \epsilon_-}\right],
\label{eq:unitary_g}
\end{equation}
where $\epsilon_\pm = \epsilon_{\bf p \pm q}$ and $\varepsilon_\pm = \epsilon_\pm + h_{\bf p \pm q}$. 
This, together with Eq.~\eqref{eq:gammavsg} giving the connection to the many-body scattering T-matrix,
is the key result needed for calculating the Hartree self-energy.
The appearance of the dressed Green's function $G(P)$ in the definition of the many-body scattering T-matrix implies an iterative
solution. Indeed, the present theory is fully self-consistent in the sense that only dressed Green's functions
are used both in the T-matrix and in the self-energy. Fortunately, due to the simplicity of the Green's function ansatz and because
of a rapid convergence of the iteration, the resulting numerical task is very easy to solve. As will be described below,
I will also consider an alternative scheme in which the imaginary part of the self-energy is neglected.

\begin{figure}
\centering
\includegraphics[width=0.90\textwidth]{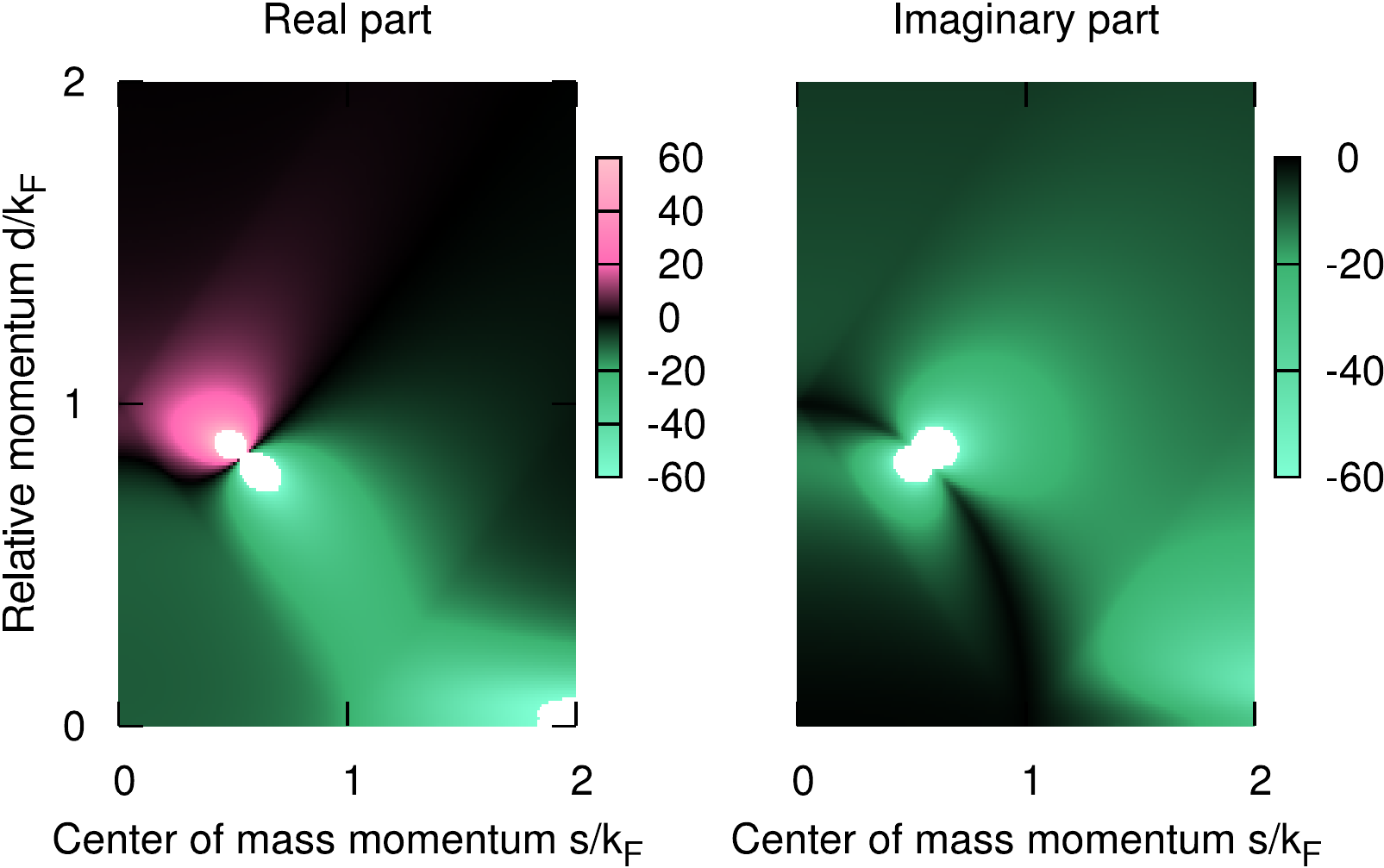}
\caption{The real and imaginary parts of $\Gamma (s,d)$ as a function of the center-of-mass momentum $s$ and the relative momentum $d$ calculated
for a zero-temperature Fermi sphere.}
\label{fig:t_sd_t0}
\end{figure}

\begin{figure}
\centering
\includegraphics[width=0.90\textwidth]{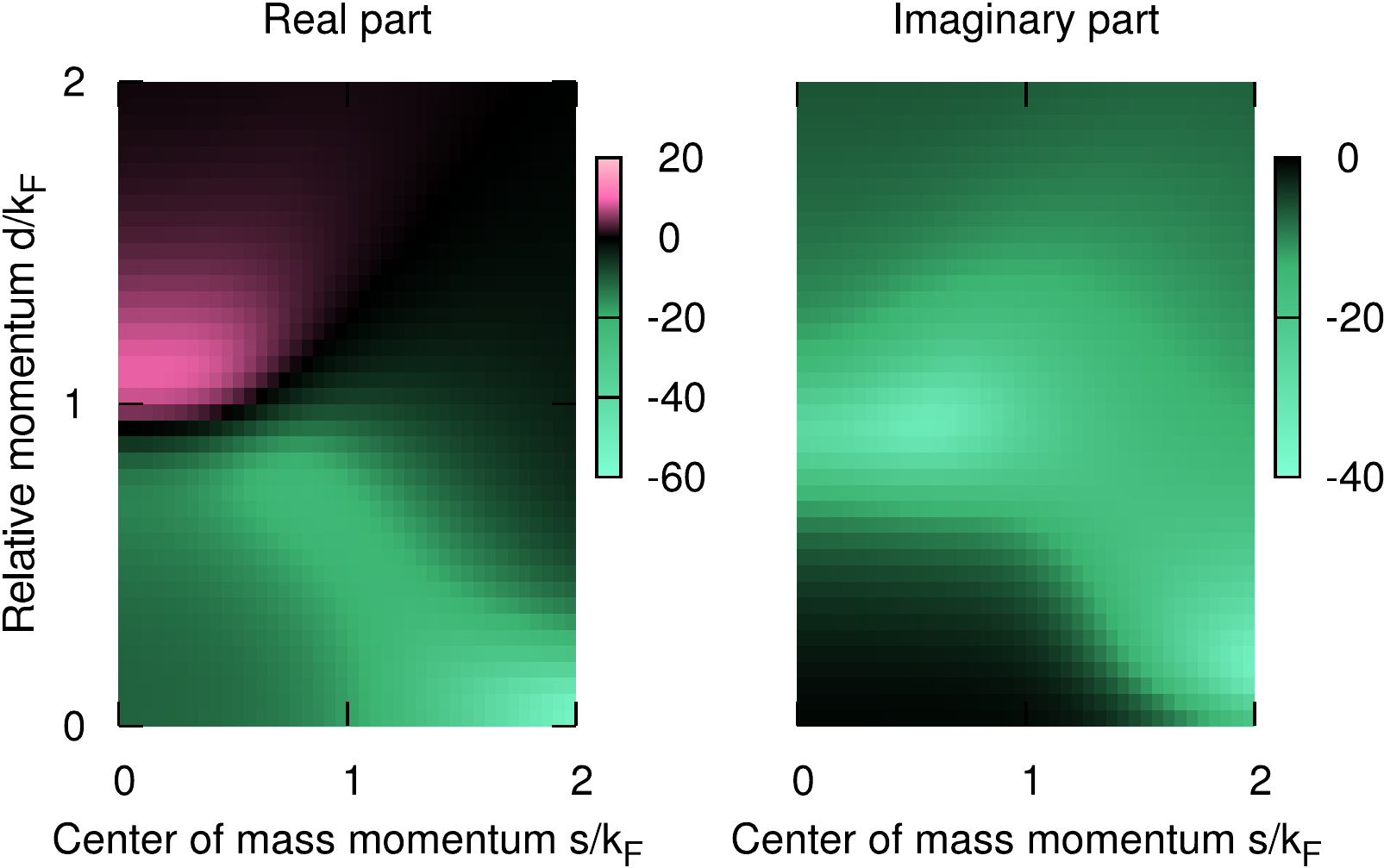}
\caption{The real and imaginary parts of $\Gamma (s,d)$ as a function of the center-of-mass momentum $s$ and the relative momentum $d$ calculated
for a finite temperature $T = 0.5\,T_\mathrm{F}$ Fermi sphere.}
\label{fig:t_sd_t05}
\end{figure}

\begin{figure}
\centering
\includegraphics[width=0.90\textwidth]{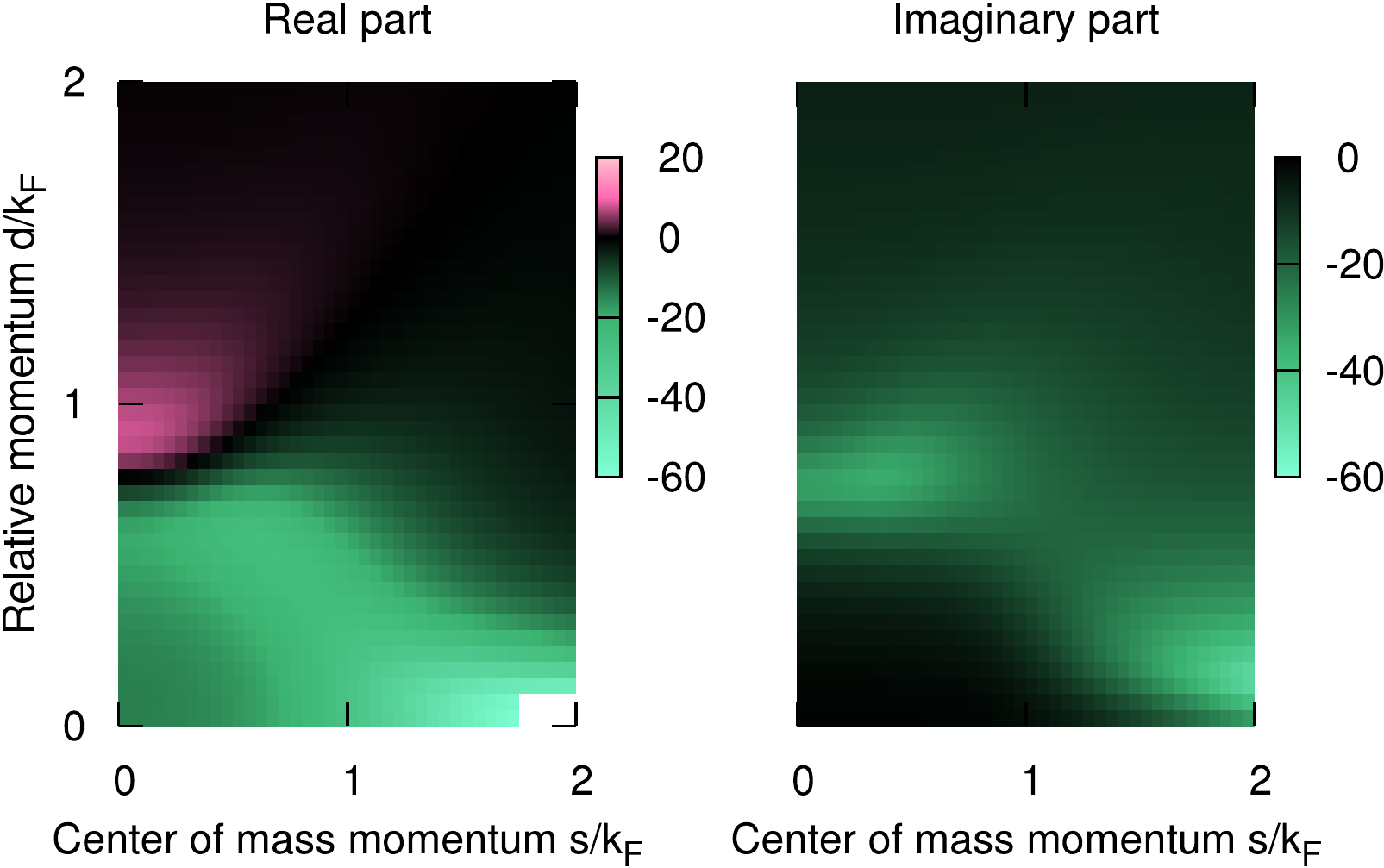}
\caption{The real and imaginary parts of $\Gamma (s,d)$ as a function of the center-of-mass momentum $s$ and relative momentum $d$ calculated
for the BPC ansatz occupation probabilities $n(x) = \frac{1}{2} \left( 1-\frac{x-\mu}{\sqrt{(x-\mu)^2 + \Delta^2}}\right)$ with
$\mu = 0.59\,E_\mathrm{F}$ and $\Delta = 0.69\, E_\mathrm{F}$.}
\label{fig:t_sd_bcs}
\end{figure}

\subsection{Numerical solution of the many-body T-matrix}

In order to obtain a better physical picture of how the many-body scattering
T-matrix looks like, I will for the moment neglect the Hartree self-energies $h(p)$ in 
Eq.~\eqref{eq:unitary_g}. As will be seen below, for the determination
of the Hartree self-energy, it is sufficient to consider elastic scatterings for
which $P = (\frac{\bf p+k}{2},\frac{\epsilon_p + \epsilon_k}{2})$. The
many-body T-matrix becomes now a function of only two parameters, the magnitudes 
of the center-of-mass momentum $s$ and the relative momentum $d$. 

Fig.~\ref{fig:t_sd_t0}
shows the real and imaginary parts of the many-body T-matrix $\Gamma (s,d)$.
The most striking feature is the divergence of the real part for small relative
momenta $d$ at sufficiently high center of mass momenta $s > k_\mathrm{F}$. This is caused by
the $1/d$ divergence of the two-body scattering T-matrix $\Gamma_0$ but
one should remember that $\Gamma_0$ was a purely imaginary quantity for elastic scatterings, 
whereas also the real part of $\Gamma$ diverges. The imaginary part of $\Gamma$ reproduces this behavior too, 
but one should also notice the suppression of the divergence for low center-of-mass momenta $s$. 
There is also a peculiar feature appearing around $d = 0.849\,k_\mathrm{F}$, $s = 0.546\,k_\mathrm{F}$. 
The physics of this feature will become clear shortly.

\begin{figure}
\centering
\includegraphics[width=0.90\textwidth]{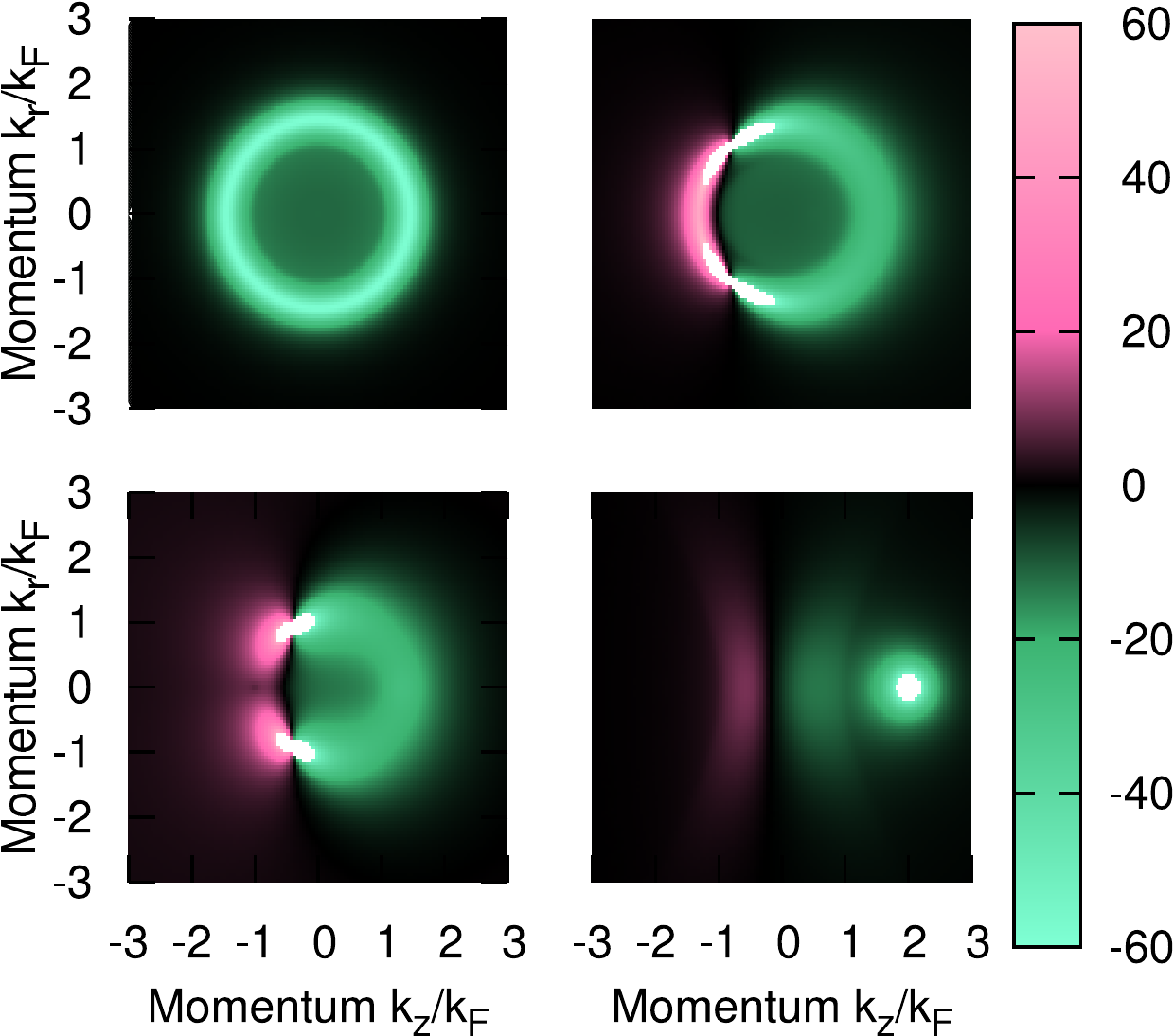}
\caption{The real part of $\Gamma (k_z,k_r)$ as a function of the atom momentum ${\bf k} = (k_z,k_r)$ for an atom with momenta ${\bf p} = 0$ (top-left), ${\bf p} = 0.50\,k_\mathrm{F}$ (top-right), ${\bf p} = 1.0\,k_\mathrm{F}$ (bottom-left), and ${\bf p} = 2.0\,k_\mathrm{F}$ (bottom-right). Shown is the absolute value of the
real part, calculated assuming the zero-temperature Fermi distribution for the occupation numbers.}
\label{fig:t_f10_p025}
\end{figure}

\begin{figure}
\centering
\includegraphics[width=0.90\textwidth]{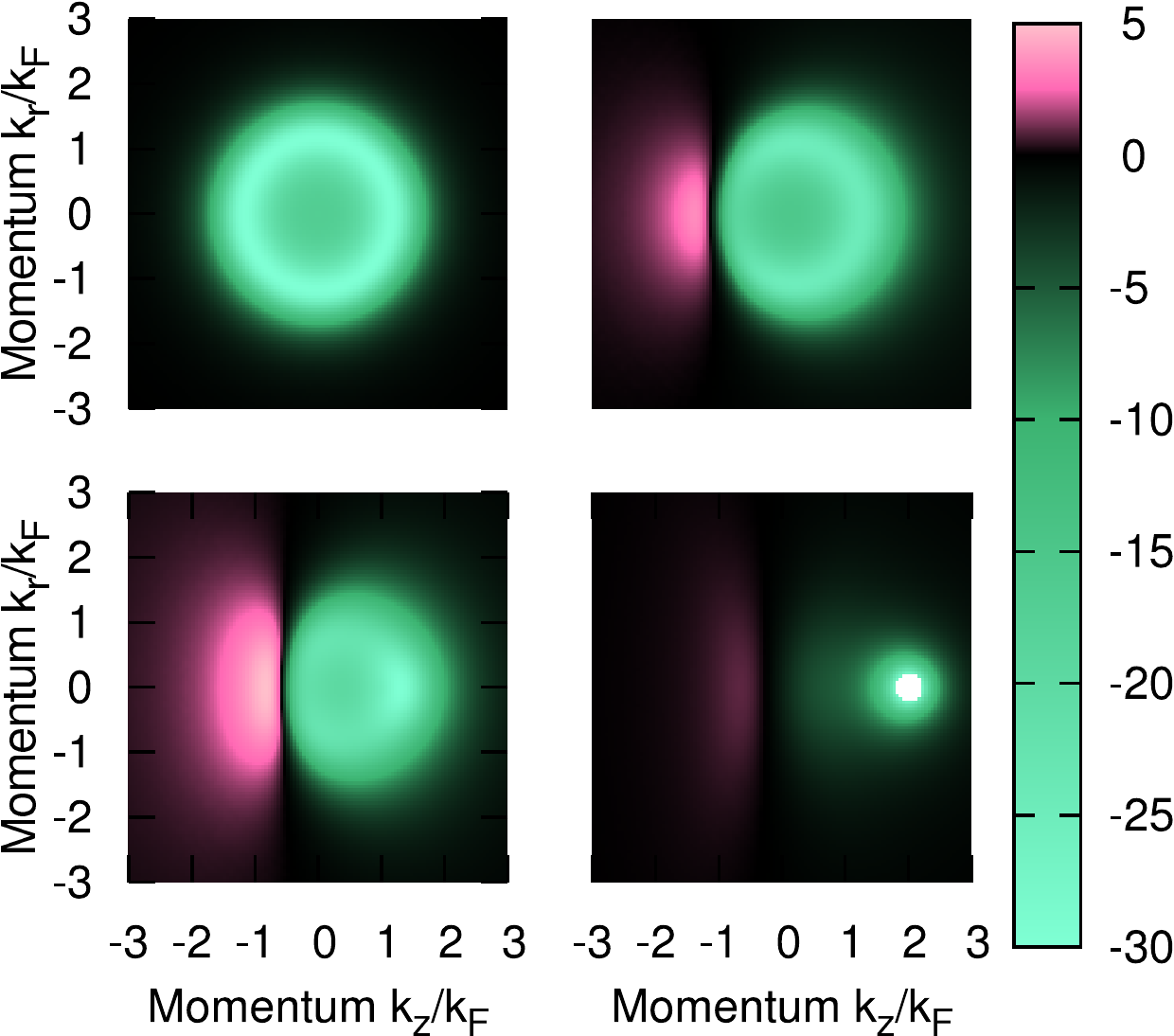}
\caption{The real part of $\Gamma (k_z,k_r)$ as a function of the atom momentum ${\bf k} = (k_z,k_r)$ for an atom with momenta ${\bf p} = 0$ (top-left), ${\bf p} = 0.50\,k_\mathrm{F}$ (top-right), ${\bf p} = 1.0\,k_\mathrm{F}$ (bottom-left), and ${\bf p} = 2.0\,k_\mathrm{F}$ (bottom-right). Shown is the absolute value of the
real part, calculated assuming the BPC ansatz for the occupation numbers.}
\label{fig:t_bcs_p025}
\end{figure}

Fig.~\ref{fig:t_sd_t05} shows the many-body T-matrix for 
a high temperature $T = 0.5\,T_\mathrm{F}$ Fermi sphere, showing
the broadening of the features when the Fermi surface becomes
less well defined. In addition, Fig.~\ref{fig:t_sd_bcs} shows the 
many-body T-matrix calculated assuming
the BPC ansatz in Eq.~\eqref{eq:bpcansatz} for the occupation numbers
$n(x) = \frac{1}{2} \left( 1-\frac{x-\mu}{\sqrt{(x-\mu)^2 + \Delta^2}}\right)$ 
with $\mu = 0.59\,E_\mathrm{F}$ and $\Delta = 0.69\, E_\mathrm{F}$.
Because of the similarity between the results obtained using this ansatz 
and the high temperature ansatz I will below concentrate only on two cases: 
the zero temperature (with a well defined Fermi surface) and the BPC ansatz. 

Despite the simplicity of the $\Gamma (s,d)$ plots, it is still rather hard to see
from these plots which interaction channels are actually dominating. 
Fig.~\ref{fig:t_f10_p025} shows the many-body scattering T-matrix as
a function of actual atom momenta ${\bf k}$ for a test atom with a momentum ${\bf p}$,
shown for different values of $p$. The strong peak observed for an atom with high momentum
$p$ is easily understood as arising from the $1/d$ divergence in the 
high center-of-mass momentum scatterings. In contrast, the scattering 
channels for a low momentum atom ($p = 0\,k_\mathrm{F}$) show a very
distinct ring with a radius of roughly $k=\sqrt{2}\,k_\mathrm{F}$. 
This ring corresponds to the small feature seen at the low center-of-mass 
momentum in Fig.~\ref{fig:t_sd_t0}. The radius of this ring, and the
corresponding remnants of the ring for intermediate momenta ($p = 0.5\,k_\mathrm{F}$),
suggest that the atoms would interact most strongly with such atoms that allow
on-shell scatterings onto the Fermi surface. Indeed, two atoms
with momenta $p=0$ and $k = \sqrt{2}\,k_\mathrm{F}$ can scatter 
right on to the Fermi surface with magnitudes of momenta 
$q = k_\mathrm{F}$. Likewise, an atom with momentum 
$p = 0.50\,k_\mathrm{F}$ has a strong interaction peak with an atom with
momentum ${\bf k} \approx (0.805,1.05)\,k_\mathrm{F}$. The total kinetic energy of these
two atoms equals $2\,E_\mathrm{F}$, allowing them to scatter on to the
Fermi surface. However, interference effects from different scattering
channels play an important role resulting in the appearance of clear maxima 
and minima on and inside the ring. One should notice, however, that
the many-body scattering T-matrix simply tells how strongly the atoms
with momenta ${\bf p}$,${\bf k}$ \emph{would} interact regardless of 
whether the two atoms exist or not. Indeed, for an atom 
inside the Fermi sphere, the strongest interaction channel is always 
with a momentum state \emph{outside} the Fermi sphere.

Broadening the Fermi surface by either a finite temperature or by the BPC ansatz 
makes these features smoother as seen in Fig.~\ref{fig:t_bcs_p025}. For low momenta $p$, 
a clear ring is still seen, with the radius $k \approx 0.77\,k_\mathrm{F}$ corresponding 
to the kinetic energy $\hbar^2 k^2/2m \approx 2\mu$. The imaginary parts of the many-body 
scattering T-matrix look qualitatively similar to the real parts and thus they are not plotted here.

\section{Hartree self-energy}
\label{sec:hartree}

The main quantity of interest in this work is the Hartree self-energy. As discussed in 
Section~\ref{sec:scattering} this requires a self-consistent iterative
solution and hence the inclusion of Hartree self-energies in the many-body scattering 
T-matrix in Eq.~\eqref{eq:unitary_g}. 
The self-energy was defined in Section~\ref{sec:hartree_initial} as
\begin{eqnarray}
    \Sigma (P) &= -i \int \frac{dK}{(2\pi)^4} \, G(K) \Gamma (K,P;P,K) \nonumber \\
&= -i \int \frac{dK}{(2\pi)^4} \, G({\bf k},k_0) \Gamma ({\bf d}, {\bf d}; S),
\end{eqnarray}
where $S = \left({\bf s},\frac{p_0 + k_0}{2}\right)$, ${\bf d} = \frac{1}{2} ( {\bf p} - {\bf k})$, and  ${\bf s} = \frac{1}{2} ( {\bf p} + {\bf k})$. 
Doing the $k_0$ frequency integral yields contributions from the pole
of the Green's function $G$ but also from possible poles in the upper half
of the complex plane in $\Gamma$. The first part was defined as the
Hartree self-energy. Since we are interested in the energy of a particle
with momentum ${\bf p}$, we will evaluate the self-energy at frequency
$p_0 = \varepsilon_p$. Notice that this results in an implicit equation for the Hartree self-energy, as the point where the value of the self-energy
is to be evaluated depends on the value of the self-energy through $\varepsilon_p = \epsilon_p + h_p$.

We can now write the Hartree self-energy as
\begin{equation}
    \Sigma_\mathrm{Hartree} ({\bf p},\varepsilon_p) \equiv \int \frac{d{\bf k}}{(2\pi)^3} \, n(\epsilon_k) \Gamma ({\bf d}, {\bf d}; {\bf s},\frac{\varepsilon_p+\varepsilon_k}{2})
\label{eq:hartree}
\end{equation}
Notice that only the particle branch of the Green's function $G(K)$ is relevant as the pole
corresponding to the hole branch is on the wrong side of the complex plane.
The Hartree self-energy is a complex quantity with the real part describing
the actual Hartree energy shift and the inverse of the imaginary part yielding 
the lifetime of the corresponding excitation (hole or particle excitation).
However, as will be seen below, omitting the pairing part of the self-energy
causes problems with the imaginary part of the total self-energy
namely that the imaginary part of $\Sigma_\mathrm{Hartree}$ is always
negative. This means that the poles corresponding to \emph{both} particle and hole excitations
are shifted upwards in the complex plane in violation of the
Lehmann representation which states that the imaginary part of the self-energy
must change sign at the Fermi momentum~\cite{FetterAndWalecka}. 


This problem however does not manifest in the present model because of the choice
of the Green's function ansatz in Eq.~\eqref{eq:green} where the imaginary part
of the self-energy in the hole and particle branches have opposite signs by design.
The correct behavior of the poles is thus forced, and since the hole and
particle branches also use the same self-energy the problem does not appear.
Being aware of these problems with the imaginary parts of the self-energy, I will
also consider an alternative path by relaxing the requirement of self-consistency
and omit the imaginary parts of the self-energy.
Such approach certainly underestimates the effect of lifetime broadening (by neglecting it altogether)
while the fully self-consistent approach most likely overestimates it since the imaginary part of the
self-energy should vanish at the Fermi surface~\cite{Luttinger1961a} but it does not
in the present model. We will use the self-consistent approach in most of the results below
because the non-zero imaginary part of the self-energy broadens spectral peaks making
numerical integrations easier. Both approaches give qualitatively similar
results and they can probably be considered as the upper and lower bounds for the Hartree
energy shifts and the excitation lifetimes.

\subsection{Weakly interacting limit}

The problems with the imaginary part of the self-energy can be easily observed
by first considering the weakly interacting (and low momentum) limit.
This calculation is effectively a review
of the Galitskii's calculation for the lifetime and energy shift of 
quasiparticle excitations.

For weakly interacting atoms the scattering amplitude can be approximated as
\[
   f({\bf s},{\bf k}) = \approx -a + ika^2,
\]
yielding the two-body scattering T-matrix
\begin{eqnarray}
\fl    \Gamma_0({\bf s}, {\bf k}; P) &= \frac{4\pi \hbar^2 a}{m} - i\frac{4\pi \hbar^2}{m} ka^2 + 
\left( \frac{4\pi \hbar^2 a}{m} \right)^2 \int \frac{d{\bf q}}{(2\pi)^3} \left( \frac{1}{2p_0 -2\epsilon_p - 2\epsilon_q} + \frac{1}{2\epsilon_q - 2\epsilon_{k} - i\eta} \right) \nonumber \\
\fl &= \frac{4\pi \hbar^2 a}{m} + \left( \frac{4\pi \hbar^2 a}{m} \right)^2 \int \frac{d{\bf q}}{(2\pi)^3} \left( \frac{1}{2p_0 -2\epsilon_p - 2\epsilon_q} + \frac{\mathcal{P}}{2\epsilon_q - 2\epsilon_k} \right),
\end{eqnarray}
where $\mathcal{P}$ signifies the principal value integration and terms of order $\mathcal{O} (a^3)$
have been dropped out. Notice that this does not depend on ${\bf s}$ but it \emph{does} depend on ${\bf k}$.

Assuming ideal zero-temperature Green's functions $G(K) = G_0(K)$ for the intermediate scattering
states, the many-body scattering T-matrix becomes (here too omitting terms $\mathcal{O} (a^3)$)
\begin{eqnarray}
\fl \Gamma (&{\bf s},{\bf k};P) = \Gamma_0 ({\bf s},{\bf k};P) + \int \frac{d{\bf q}}{(2\pi)^3} \, \Gamma_0 ({\bf s},{\bf q};P) \\
\fl &\quad \quad \times \left[ \frac{n(-\epsilon_{{\bf p} + {\bf q}})n(-\epsilon_{{\bf p} - {\bf q}})-1}{2p_0 - 2\epsilon_p - 2\epsilon_q} - \frac{n(\epsilon_{{\bf p} + {\bf q}}) n(\epsilon_{{\bf p} - {\bf q}})}{2p_0^* - 2\epsilon_p - 2\epsilon_q} \right]\Gamma ({\bf q}, {\bf k};P) \nonumber \\
\fl &\approx \frac{4\pi \hbar^2 a}{m} + \left( \frac{4\pi \hbar^2 a}{m} \right)^2 \int \frac{d{\bf q}}{(2\pi)^3} \left( \frac{1}{2p_0 -2\epsilon_p - 2\epsilon_q} + \frac{\mathcal{P}}{2\epsilon_q - 2\epsilon_k} \right) \nonumber \\
\fl &\quad \quad+ \int \frac{d{\bf q}}{(2\pi)^3} \, \frac{4\pi \hbar^2 a}{m} \left[ \frac{n(-\epsilon_{{\bf p} + {\bf q}})n(-\epsilon_{{\bf p} - {\bf q}})-1}{2p_0 - 2\epsilon_p - 2\epsilon_q} - \frac{n(\epsilon_{{\bf p} + {\bf q}}) n(\epsilon_{{\bf p} - {\bf q}})}{2p_0^* - 2\epsilon_p - 2\epsilon_q} \right] \frac{4\pi \hbar^2 a}{m} \nonumber \\
\fl &= \frac{4\pi \hbar^2 a}{m} + \left( \frac{4\pi \hbar^2 a}{m} \right)^2 \int \frac{d{\bf q}}{(2\pi)^3} \left( \frac{\mathcal{P}}{2\epsilon_q - 2\epsilon_{k}} + \frac{n(-\epsilon_{{\bf p} + {\bf q}})n(-\epsilon_{{\bf p} - {\bf q}})}{2p_0 - 2\epsilon_p - 2\epsilon_q} - \frac{n(\epsilon_{{\bf p} + {\bf q}}) n(\epsilon_{{\bf p} - {\bf q}})}{2p_0^* - 2\epsilon_p - 2\epsilon_q}\right), \nonumber
\end{eqnarray}
and the Hartree self-energy is now
\begin{eqnarray}
\fl  \Sigma_\mathrm{Hartree} &= \int \frac{d{\bf k}}{(2\pi)^3} \, n(\epsilon_k) \Gamma \left({\bf d},{\bf d};{\bf s},\frac{\epsilon_p + \epsilon_k}{2} +i\eta\right) \\
\fl &= \frac{\hbar^2 k_\mathrm{F}^2}{2m} \frac{4}{3\pi} k_\mathrm{F} a + 32 \pi^2 \frac{\hbar^2 k_\mathrm{F}^2}{2m} (k_\mathrm{F} a)^2 \int \frac{d{\bf k}}{(2\pi)^3}\int \frac{d{\bf q}}{(2\pi)^3}  \theta(1-k) \nonumber \\
\fl &\times \left( \frac{\theta(|{\bf s} + {\bf q}|-1)\theta(|{\bf s} - {\bf q}|-1)}{d^2 - q^2 + i\eta'} - \frac{\theta(1-|{\bf s} + {\bf q}|) \theta(1-|{\bf s} - {\bf q}|)}{d^2 - q^2 - i\eta'} - \frac{\mathcal{P}}{d^2 - q^2}\right),\nonumber
\label{eq:weakhartree2}
\end{eqnarray}
where $\eta' = m\eta/\hbar^2$ and the relation $n(\epsilon_k) = \theta(E_\mathrm{F}-\epsilon_k) = \theta (1-k/k_\mathrm{F})$ 
has been used in order to allow a direct comparison with the results in Ref.~\cite{FetterAndWalecka}. 
The dominant (first) term is the standard weakly interacting
limit Hartree shift and the rest of the terms are corrections to this
weak interaction result.
However, a careful comparison with Ref.~\cite{FetterAndWalecka} shows that the result is missing terms 
proportional to $\theta(k-1)$ and it has an extra term proportional to $\theta (1-k)$. 
The reason for the discrepancy is the neglect of the pole in the T-matrix when 
doing the $k_0$ integration. However, as discussed in Section~\ref{sec:hartree_initial},
such effects do not enter the Hartree part of the self-energy but instead would 
go in to the pair part $\Sigma_\mathrm{pairs}$. 
Notice, in particular, that the imaginary part  of the Hartree self-energy 
in Eq.~\eqref{eq:weakhartree2} does \emph{not} change sign at the Fermi surface and
hence the imaginary part of the Hartree self-energy for hole and particle excitations would have 
the same sign. The Green's function ansatz in Eq.~\eqref{eq:green} fixes this problem
by demanding that the self-energy for the hole branch is obtained by taking the complex
conjugate of the particle branch self-energy.

\subsubsection{Unitarity}
\label{sec:Hartree}

The Hartree self-energy at unitarity can be calculated from Eq.~\eqref{eq:hartree}
\begin{eqnarray}
    \Sigma_\mathrm{Hartree} (P) &= \int \frac{d{\bf k}}{(2\pi)^3} n(\epsilon_k) \Gamma \left({\bf d}, {\bf d}; {\bf s},\frac{\varepsilon_p + \varepsilon_k}{2} \right) \nonumber \\
&= \int \frac{d{\bf k}}{(2\pi)^3} n(\epsilon_k) \Gamma \left({\bf s},\frac{\varepsilon_p + \varepsilon_k}{2} \right).
\end{eqnarray}
It is instructive to consider first the case in which the many-body corrections are ignored. This can be achieved by approximating $\Gamma \approx \Gamma_0$
and $\varepsilon_p \approx \epsilon_p$, yielding
\begin{eqnarray}
\fl    \Sigma_\mathrm{Hartree}^0 (P) &=\int \frac{d{\bf k}}{(2\pi)^3} n(\epsilon_k) \frac{4\pi\hbar^2}{m} \sqrt{\frac{\hbar^2}{2m|\epsilon_{\frac{1}{2}\left( {\bf p} + {\bf k}\right)} - \frac{1}{2} \left(\epsilon_p + \epsilon_k \right)|}} \nonumber \\
\fl &\quad \quad \times \left[ \theta \left(\epsilon_{\frac{1}{2}\left( {\bf p} + {\bf k}\right)} - \frac{1}{2} \left(\epsilon_p + \epsilon_k \right) \right) - i \theta \left(\frac{1}{2} \left(\epsilon_p + \epsilon_k \right) - \epsilon_{\frac{1}{2}\left( {\bf p} + {\bf k}\right)}\right) \right] \nonumber \\
\fl &=-i\frac{4\pi\hbar^2}{m} \int \frac{d{\bf k}}{(2\pi)^3} n(\epsilon_k) \frac{2}{\left|{\bf p} - {\bf k}\right|}.
\end{eqnarray}
The lesson here is that the Hartree self-energy does indeed involve the relative momentum ${\bf p}-{\bf k}$ instead of the center-of-mass momentum. 
This is a result of the way the center of mass quantities are formed, namely that the energy difference 
$\frac{\epsilon_p + \epsilon_k}{2} - \epsilon_s = \epsilon_d$. The same energy difference enters also the many-body T-matrix and thus also it will 
depend on the momentum difference ${\bf d}$.
Notice that the Hartree self-energy obtained using the two-body scattering T-matrix $\Gamma_0$ is purely imaginary and thus the atoms would not 
experience any energy shifts but
excitations would only have a finite lifetime. The next step is to evaluate the Hartree self-energy by including the many-body corrections.

\subsubsection{Numerical solution of Hartree self-energy at unitarity}

\begin{figure}
\centering
\includegraphics[width=0.45\textwidth]{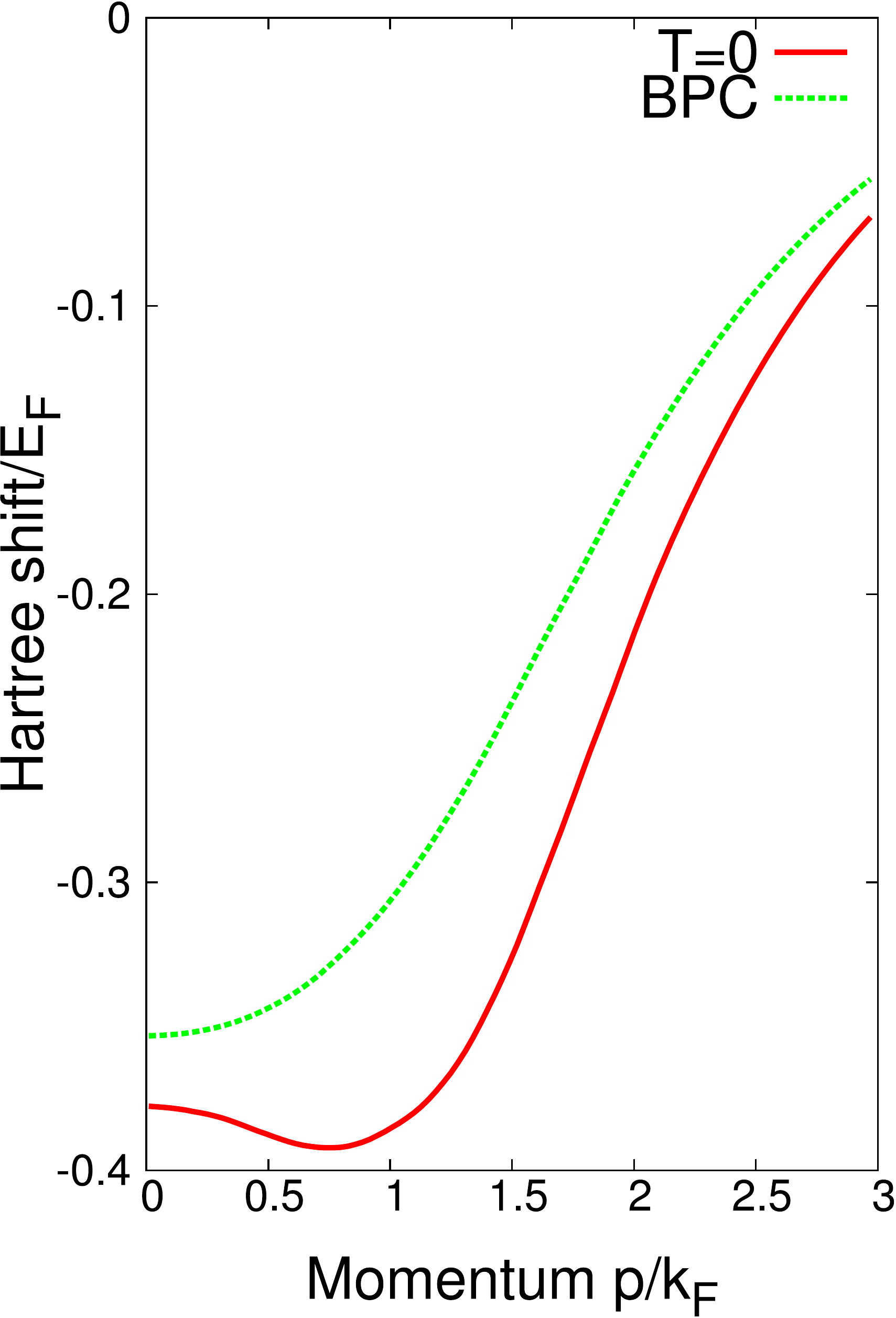}
\includegraphics[width=0.45\textwidth]{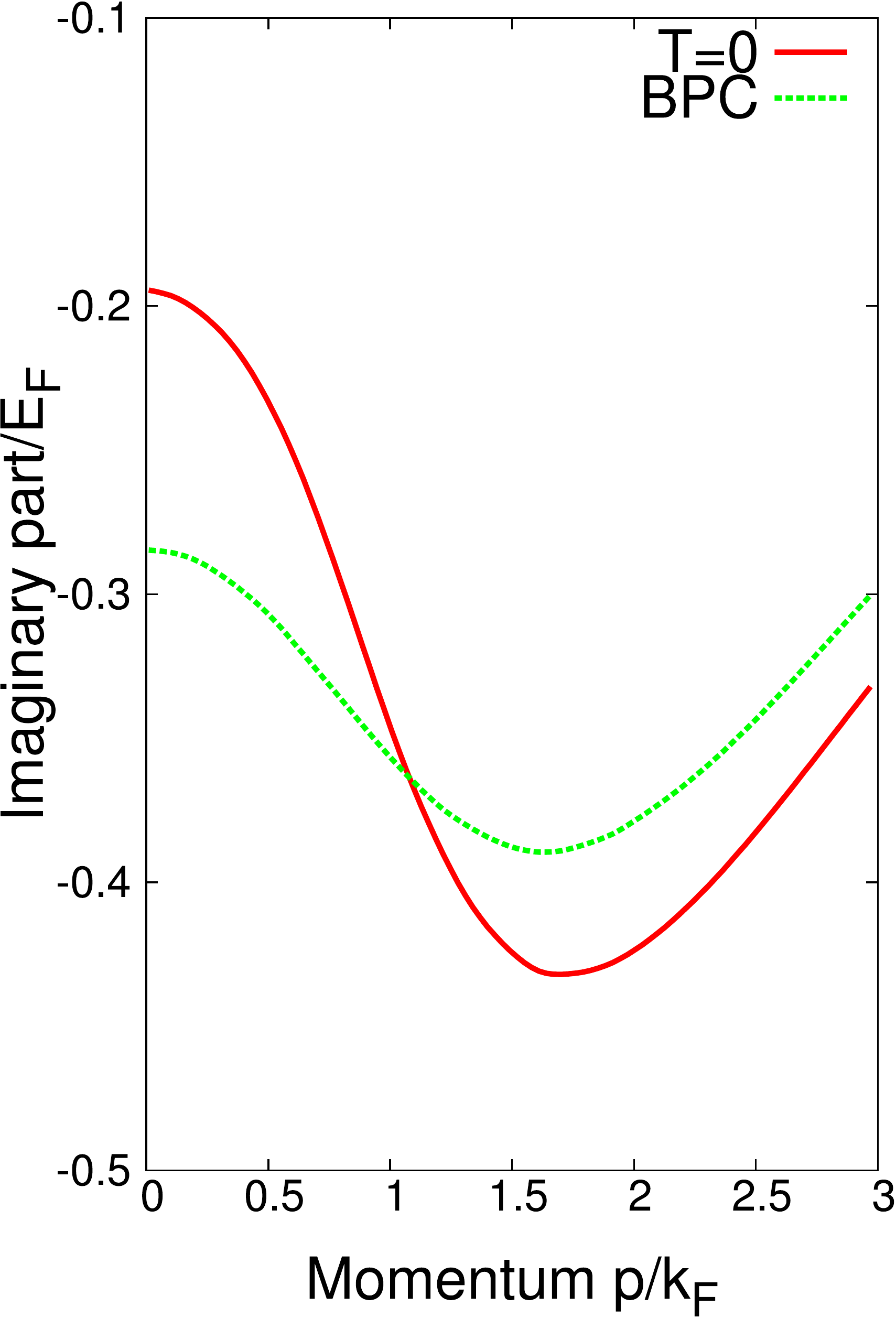}
\caption{The real and imaginary parts of the Hartree self-energy $\Sigma_\mathrm{Hartree} (p,\varepsilon_p)$ as a function of the atom
momentum $p$ for the ideal Fermi sphere and the BPC ansatz.}
\label{fig:hartree_unitary}
\end{figure}

Fig.~\ref{fig:hartree_unitary} shows the Hartree self-energy $\Sigma_\mathrm{Hartree} (P)$ 
as a function of the atom momentum $p$ (and for energy $p_0 = \varepsilon_p$) for
both the BPC ansatz occupation distribution and the zero-temperature Fermi distribution.
The Hartree shift is strongly momentum dependent, although for low momenta 
$p < k_\mathrm{F}$ the dependence is not very strong. However, beyond
the Fermi momentum, the Hartree shift exhibits a $1/p$ decay.
It is also worth noticing that the energy shift is higher at low 
momenta for the BPC ansatz than for the well-formed Fermi sphere. This is
because of the appearance of holes in the Fermi sphere, relaxing
the effect of the Pauli blocking. This effect can be seen also by increasing the temperature
in the Fermi distribution.

Shown is also the imaginary part of the Hartree self-energy with a pronounced
peak slightly above the Fermi momentum. As discussed in Section~\ref{sec:hartree}, 
this behavior is at odds with 
the Galitskii's result for a weakly interacting gas, where the
imaginary part vanishes at the Fermi momentum. On the other hand,
the pair part of the self-energy $\Sigma$ may also provide an imaginary
part that can cancel at least partly the imaginary part in the Hartree
self-energy $\Sigma_\mathrm{Hartree}$. 

\begin{figure}
\centering
\includegraphics[width=0.45\textwidth]{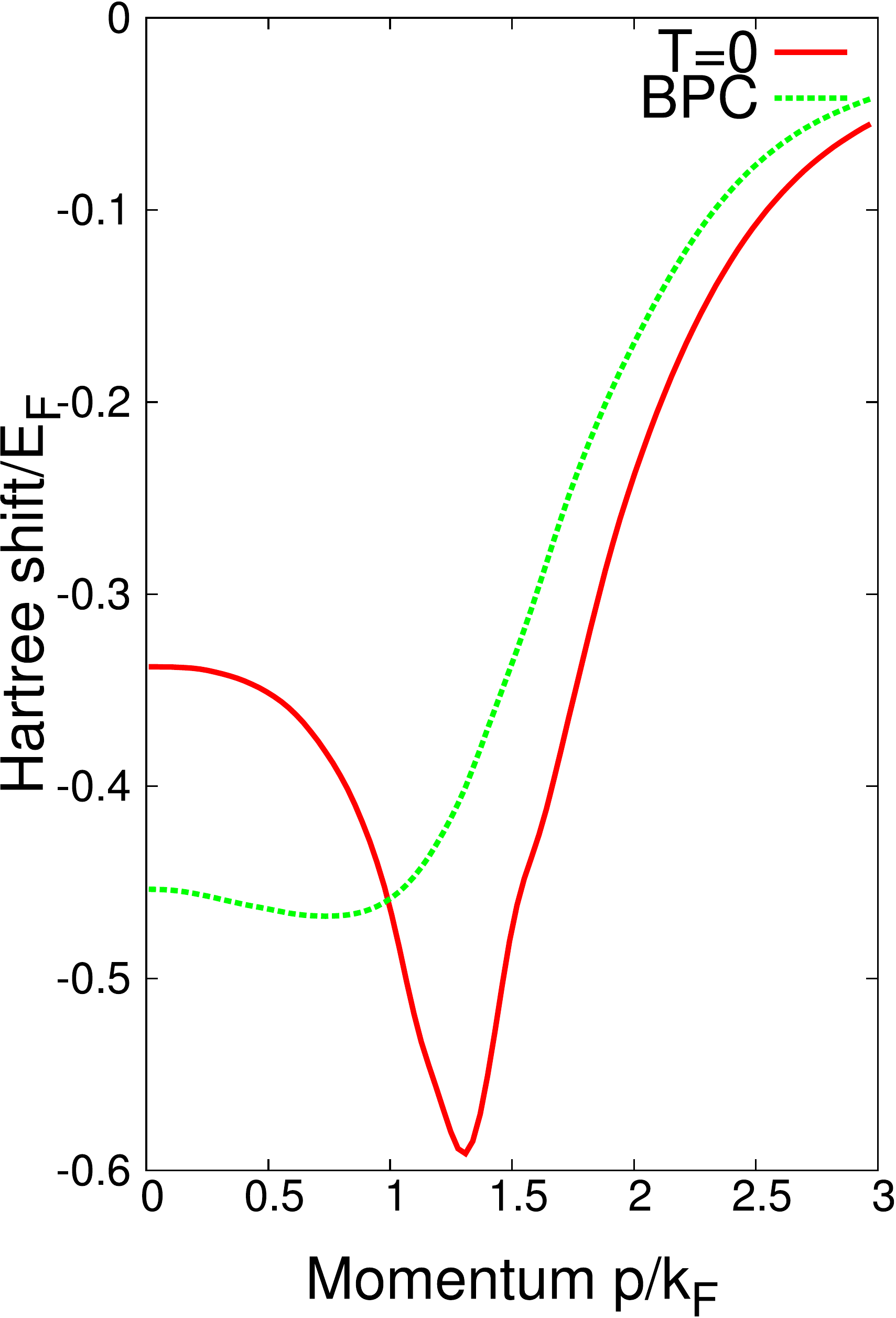}
\includegraphics[width=0.45\textwidth]{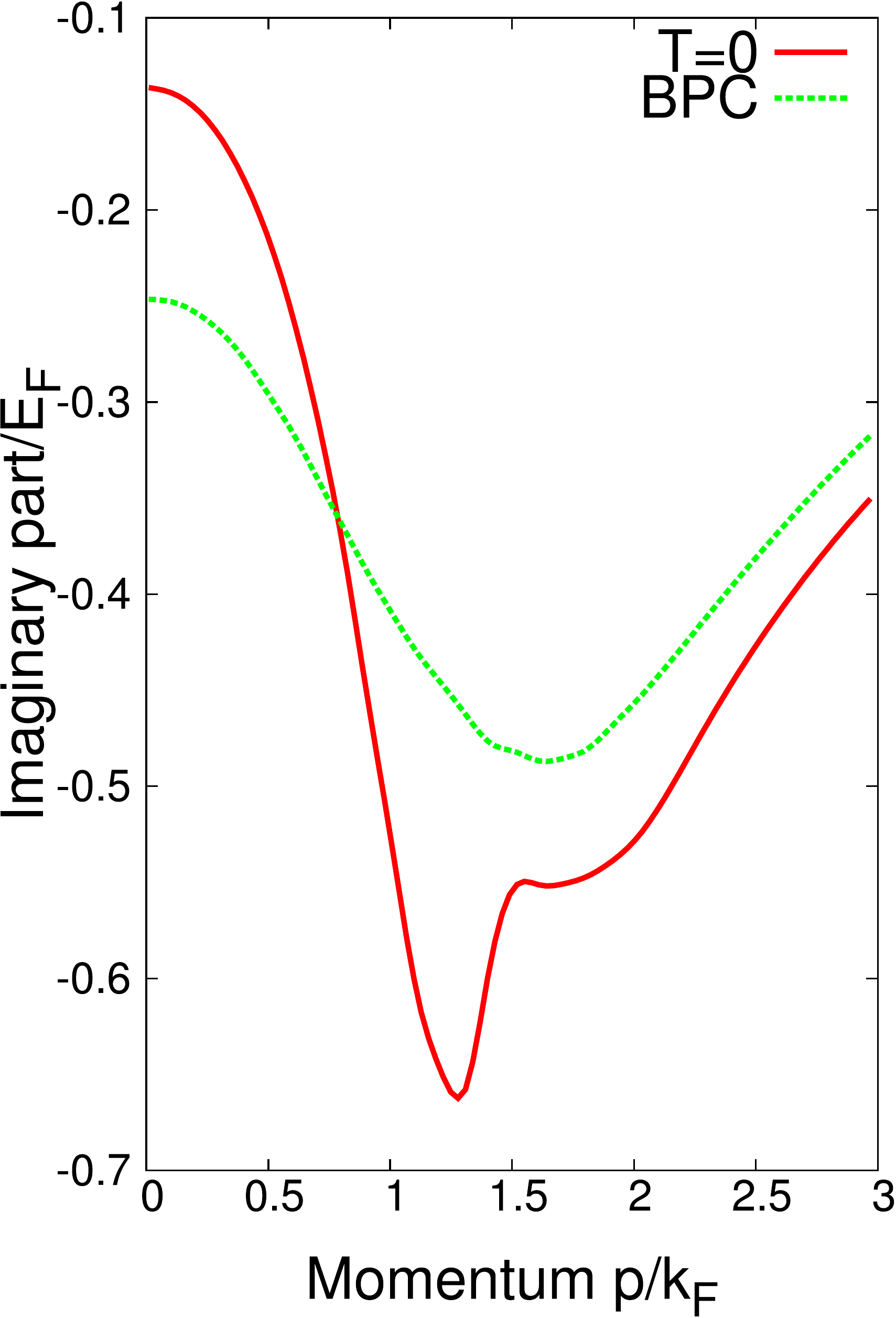}
\caption{The real and imaginary parts of the Hartree self-energy $\Sigma_\mathrm{Hartree} (p,\varepsilon_p)$ as a function of the
atom momentum $p$ for the ideal Fermi sphere and the BPC ansatz, calculated using the semi-self-consistent scheme in which 
the imaginary part of the self-energy is neglected.}
\label{fig:hartree_unitary_noimag}
\end{figure}
Fig.~\ref{fig:hartree_unitary_noimag} shows the real and imaginary
parts of the Hartree self-energy calculated using the alternative scheme in which the
imaginary part of the self-energy is omitted altogether. Notice that even though the imaginary
part of the self-energy is not included in the iteration, its value can still be calculated
and hence shown in the figure.
While the numerical values for the two schemes 
do differ somewhat, the qualitative behavior is similar for the two approaches. 
As argued in Section~\ref{sec:hartree} the 'true' behavior is expected to be 
somewhere between these two limits.

\section{Hartree self-energy in a spin-imbalanced gas}
\label{sec:hartreeforpolarized}

The present method can be easily generalized for studying polarized, or 
spin-imbalanced, gases. This can be accomplished by having spin-dependent 
distribution functions $n_\sigma (x)$ and Hartree shifts $h_{\sigma,p}$, with 
$\sigma \in \{ \uparrow, \downarrow\}$. The only equation that needs
to be modified is the $g(P)$ function in Eq.~\eqref{eq:unitary_g},
which becomes
\begin{equation}
\fl  g(P) = \int \frac{d{\bf q}}{(2\pi)^3} \left[ \frac{n_\uparrow(-\epsilon_{{\bf p} + {\bf q}})n_\downarrow(-\epsilon_{{\bf p} - {\bf q}})}{2p_0 - \varepsilon_{\uparrow,+} - \varepsilon_{\downarrow,-}} - \frac{n_\uparrow(\epsilon_{{\bf p} + {\bf q}}) n_\downarrow(\epsilon_{{\bf p} - {\bf q}})}{2p_0^* - \varepsilon_{\uparrow,+}^* - \varepsilon_{\downarrow,-}^*} - \frac{1}{2\epsilon - 2\epsilon_p - 2\epsilon_q}\right],
\end{equation}
where $\varepsilon_{\sigma,\pm} = \epsilon_\pm + h_{\sigma,\pm}$. Notice
that $g(P)$ is symmetric with respect to the change $\uparrow \longleftrightarrow \downarrow$ and 
the functions $g$ and $\Gamma$ are independent of the spin.

The Hartree self-energy Eq.~\eqref{eq:hartree} for atoms with the spin $\uparrow$ becomes
\begin{equation}
  \Sigma_\mathrm{Hartree}^\uparrow \approx \int \frac{d{\bf k}}{(2\pi)^3} \, n_\downarrow(\epsilon_k) \Gamma \left({\bf d}, {\bf d}; {\bf s},\frac{\varepsilon_{\uparrow,p} + \epsilon_{\downarrow,k}}{2} \right),
\end{equation}
and for spin $\downarrow$ atoms it becomes
\begin{equation}
  \Sigma_\mathrm{Hartree}^\downarrow \approx \int \frac{d{\bf k}}{(2\pi)^3} \, n_\uparrow(\epsilon_k) \Gamma \left({\bf d}, {\bf d}; {\bf s},\frac{\varepsilon_{\downarrow,p} + \varepsilon_{\uparrow,k}}{2} \right).
\end{equation}

\begin{figure}
\centering
\includegraphics[width=0.45\textwidth]{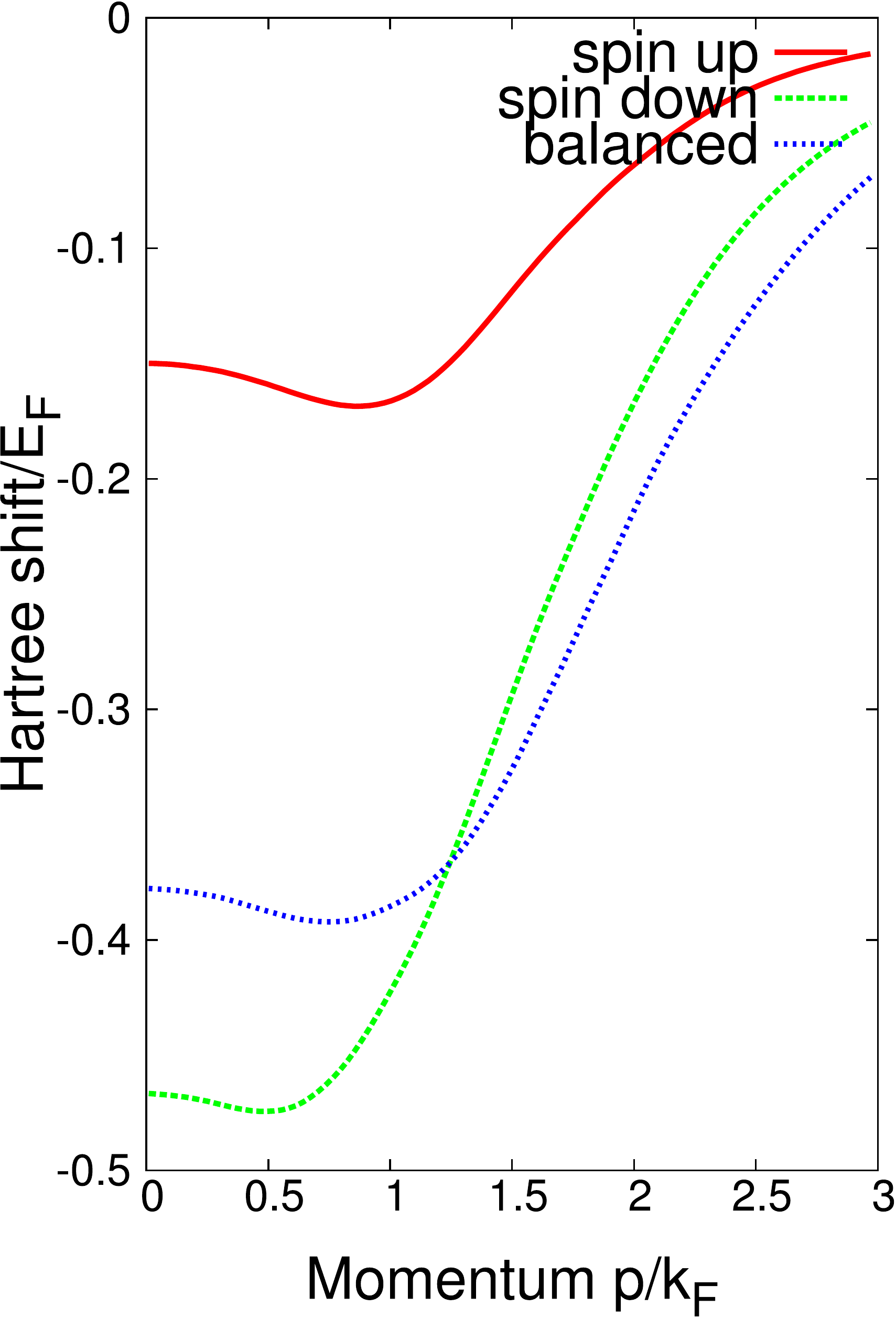}
\includegraphics[width=0.45\textwidth]{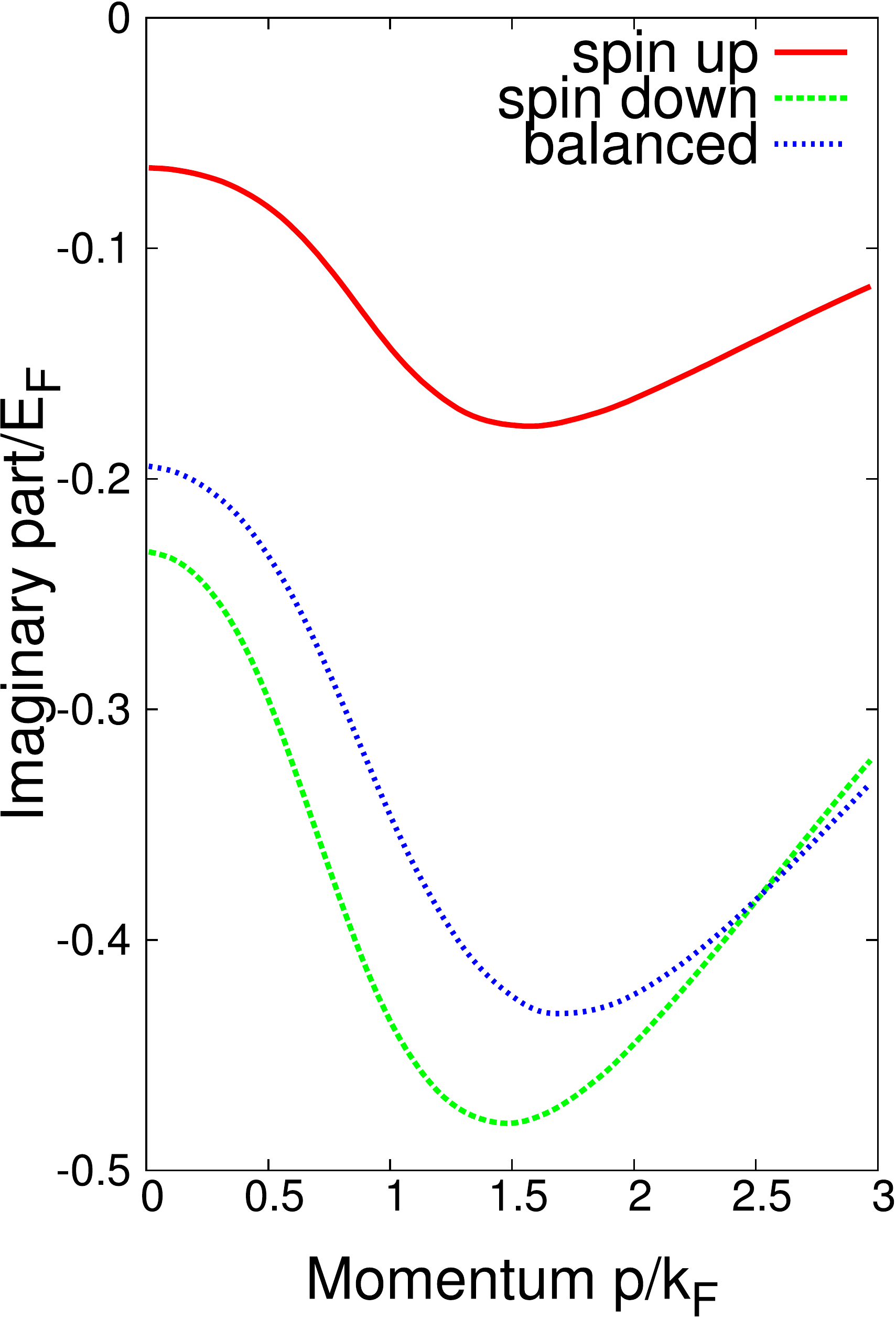}
\caption{The Hartree self-energy of a polarized ($P = 0.5$) zero-temperature Fermi gas. The real part is shown in the left and the imaginary part on the right.
The Hartree self-energy for the balanced gas is shown for reference.}
\label{fig:hartree_polar_f10_p050}
\end{figure}

Fig.~\ref{fig:hartree_polar_f10_p050} shows
the Hartree self-energy for a spin-imbalanced gas with the polarization 
$P \approx 0.5$ at the zero temperature. What is striking is the increase of the Hartree energy shift
experienced by the minority component as compared to the balanced gas (the number of majority component atoms is the same for balanced and spin-imbalanced cases). 
This is the result of the relaxation of the Pauli blocking, as the low momentum minority component atoms have access also to low momentum scatterings with 
majority component atoms close to the Fermi surface. In contrast, in a balanced gas such low momentum scatterings are blocked and 
the 'lowest momentum' scattering would amount to lifting the atom onto the surface of the Fermi sphere. Obviously the corresponding 
energy shift is decreased for the majority component as there are fewer minority component atoms to interact with.

\subsection{Polaron limit}

\begin{figure}
\centering
\includegraphics[width=0.45\textwidth]{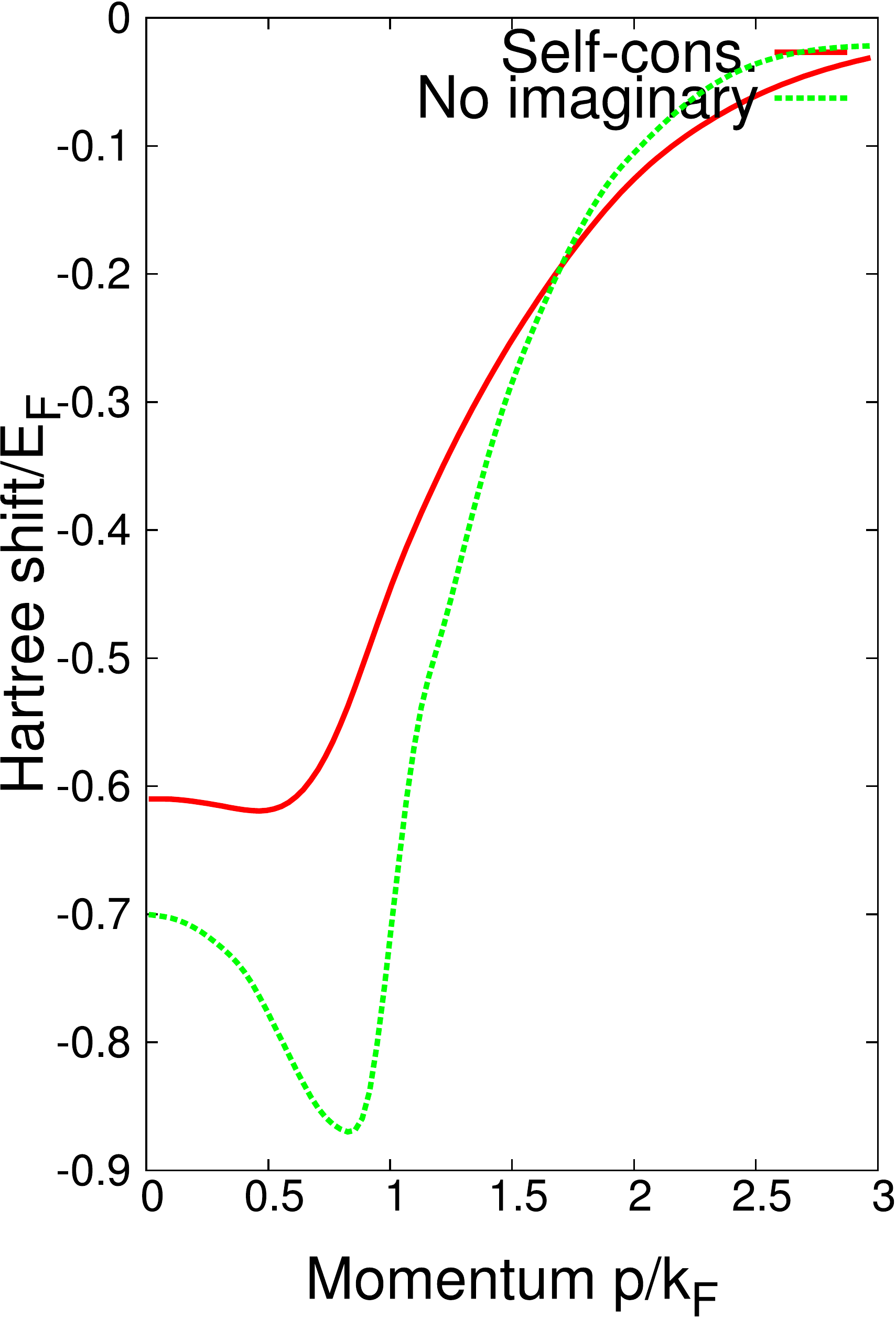}
\includegraphics[width=0.45\textwidth]{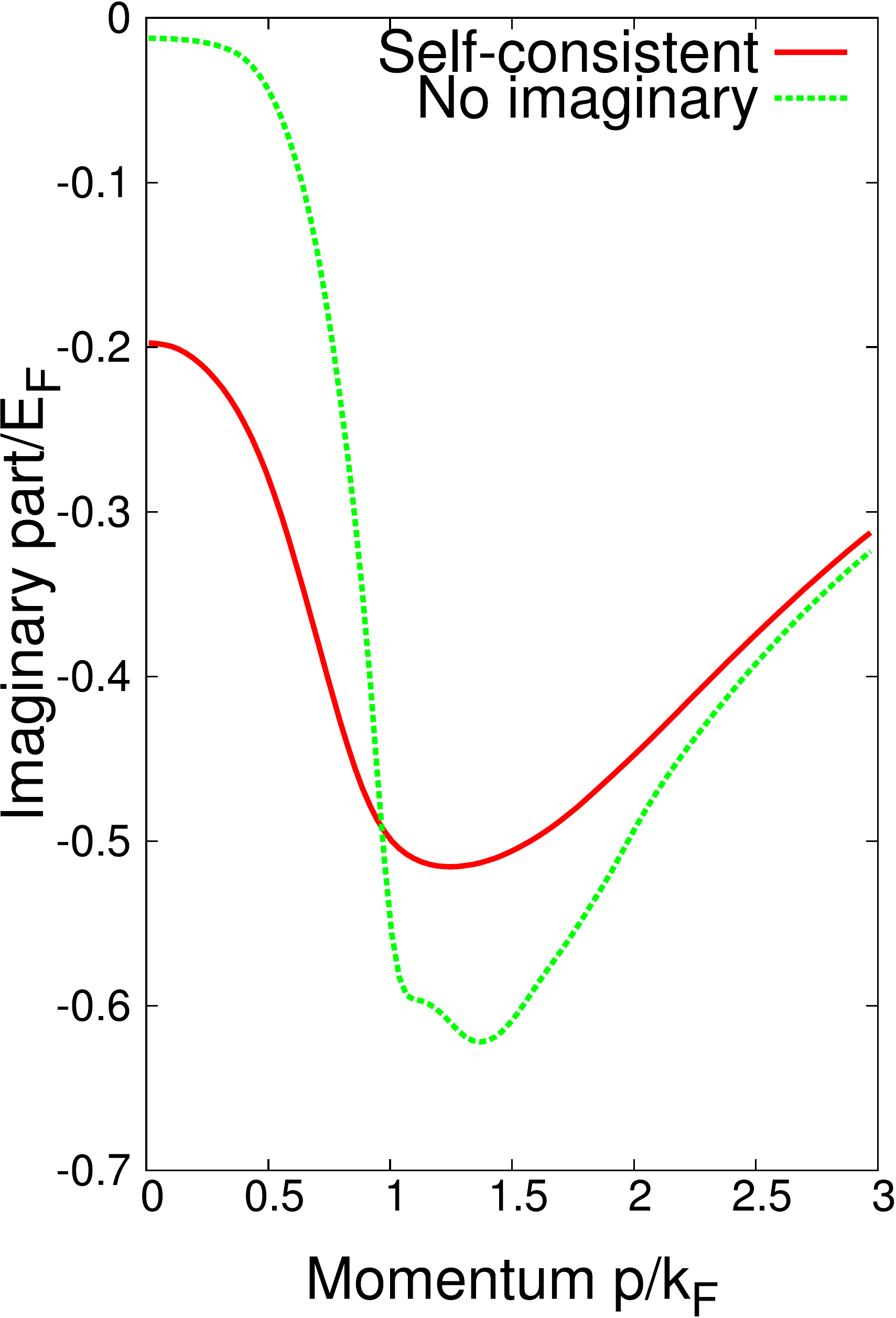}
\caption{The real and imaginary part of the Hartree self-energy of a single minority component atom immersed
in a zero-temperature Fermi sea of majority component atoms. Shown are results from both
the fully self-consistent scheme and from the semi-self-consistent scheme in which the imaginary parts
of the self-energy are neglected in the iteration.}
\label{fig:hartree_polaron}
\end{figure}

An interesting special case is the extreme polarization limit $P \rightarrow 1$ 
where one has only a single minority component atom as an impurity interacting
with an ideal Fermi sphere of majority component atoms. Such a setting is 
predicted to create polarons and it has been already widely studied both
theoretically~\cite{Lobo2006a,Chevy2006a,Combescot2007a,Prokofev2008a,Combescot2008a,Punk2009a} 
and experimentally~\cite{Shin2008a,Schirotzek2009a,Nascimbene2009a,Nascimbene2010a}. 
In the polaron limit one would expect
that the pair part of the self-energy has the least effect and the prediction
from the Hartree part would be thus of particular interest.

However, first I would like to make a connection to the polaron theory in 
Ref.~\cite{Combescot2007a}, which also led to the same analytical formula
as the variational Chevy ansatz in~\cite{Chevy2006a}. In Ref.~\cite{Combescot2007a} 
the self-consistent iteration was cut after the first iteration step, resulting
in a theory where the self-energy of a particle with momentum ${\bf p}$ was
evaluated at energy $\epsilon_p + h_{\downarrow,p}$, but the Hartree shift 
$h_{\downarrow,p}$ was not fed back into the many-body scattering T-matrix,
yielding as the polaron energy at unitarity $a \rightarrow \infty$
\begin{equation}
   E = \epsilon_{p} + \sum_{|{\bf q}| < k_\mathrm{F}} \left({\sum_{\bf k} \frac{1}{\epsilon_k+\epsilon_{\bf p+q-k}-\epsilon_q-E} -\frac{1}{2\epsilon_k}}\right)^{-1}.
\label{eq:chevyresult}
\end{equation}
Notice the implicit form of the solution, which is the result of the important assumption
that although the Hartree shift is not fed into the many-body scattering T-matrix,
it does enter in the energy $E$ at which the energy shift is evaluated at.
In the language of this work, it corresponds to using bare Green's functions $G_0$ 
in Eq.~\eqref{eq:gunitarity} but evaluating the Hartree energy shift in Eq.~\eqref{eq:hartree}
at a shifted energy $\varepsilon_p = \epsilon_p + h_p$.
While this appears like a strong assumption, the results 
(with the zero-momentum polaron energy being roughly $-0.606\,E_\mathrm{F}$)  were in very good
agreement with Monte Carlo results~\cite{Lobo2006a}.
The Hartree self-energy described in this work simplifies into Eq.~\eqref{eq:chevyresult} when doing the same 
approximations but it also allows a more rigorous analysis of the justification and the meaning of these approximations.

The simplest generalization from Eq.~\eqref{eq:chevyresult} would be to include
the Hartree shift in the many-body T-matrix but assume that it is constant, i.e. it does not depend on the momentum $h_{\downarrow,p} = h_{\downarrow}$. 
This corresponds to adding the Hartree shift to the single-particle dispersions of the spin-$\downarrow$ atoms in the second term 
in Eq.~\eqref{eq:chevyresult}, yielding
\begin{equation}
   E = \epsilon_p + \sum_{|{\bf q}| < k_\mathrm{F}} \left({\sum_{|{\bf k}| > k_\mathrm{F}} \frac{1}{\epsilon_k+\epsilon_{\bf p+q-k} + h_\downarrow -\epsilon_q-E} -\frac{1}{2\epsilon_k}}\right)^{-1}.
\label{eq:chevyresult2}
\end{equation}
Notice that only one of the energies in the second term describe a spin-$\downarrow$ atom and hence the shift $h_\downarrow$ 
appears in the denominator only once (the energy shift of majority component atoms is assumed to be zero). 
However, the second term in the right is the Hartree energy shift $h_\downarrow$, and one obtains $E = \epsilon_p + h_\downarrow$, 
which upon substitution to Eq.~\eqref{eq:chevyresult2} yields
\begin{equation}
   E = \epsilon_p + \sum_{|{\bf q}| < k_\mathrm{F}} \left({\sum_{|{\bf k}| > k_\mathrm{F}} \frac{1}{\epsilon_k+\epsilon_{\bf p+q-k} -\epsilon_q-\epsilon_p} -\frac{1}{2\epsilon_k}}\right)^{-1}.
\label{eq:chevyresult3}
\end{equation}
This corresponds to the approximation in which the Hartree self-energy in Eq.~\eqref{eq:hartree} is evaluated
at energy $\epsilon_p$. Solving this equation is actually easier than Eq.~\eqref{eq:chevyresult}, yielding
as the zero-momentum polaron energy $E \approx -0.917\,E_\mathrm{F}$. The excellent agreement with the Monte Carlo result
is thus lost.

Considering these issues from the point of view of the scattering theory presented in this work, 
the reason why the Chevy ansatz works~\cite{Chevy2006a} so well is that the
self-energy is strongly momentum dependent, so that most of the intermediate
scattering states have vanishing self-energy. Indeed, due to the short range of the
interactions and the corresponding minor importance of low energy states in the
scatterings, most of the intermediate scattering states are only weakly interacting with
the majority component Fermi sphere. This in turn limits the level of correlations needed
for an efficient description of the system and the simple model provided by the Chevy 
ansatz works well.
Contrast this with a system in which the intermediate scattering states are limited by
a low momentum cutoff (for example due to a long effective range of the interaction potential
or due to a band gap in the single-particle excitation spectrum in a deep optical lattice).
In such case approximating these intermediate states with a non-interacting Green's function
would not work as well.

Solving the self-energy in a self-consistent manner by doing the required
Hartree shift iteration yields the Hartree shift shown in Fig.~\ref{fig:hartree_polaron}. 
It shows the polaron self-energy at unitarity as a function of 
the polaron momentum $p$ and, as expected, it exhibits a strong momentum dependence
with a $1/p$ decay at high momenta. At low momenta $p < k_\mathrm{F}$, the 
energy shift is roughly constant with magnitude approximately 
$-0.61\,E_\mathrm{F}$, which is in good agreement with predictions from other
theories~\cite{Lobo2006a,Chevy2006a,Combescot2007a,Prokofev2008a,Pilati2008a,Combescot2008a} 
and experiments~\cite{Shin2008a,Schirotzek2009a,Nascimbene2009a,Nascimbene2010a}.

From the momentum dependence of the self-energies one can also obtain the 
effective mass $m^*$ for the polaron~\cite{Combescot2007a}
\begin{equation}
  m^* = m\frac{1-\frac{\partial \Sigma}{\partial \omega}}{1+\frac{\partial \Sigma}{\partial \left(p^2/2m\right)}}.
\end{equation}
Ignoring the possible frequency dependence of the self-energy, I obtain
the effective mass $m^* \approx 1.04\,m$ at unitarity, which agrees well with 
Monte Carlo results~\cite{Lobo2006a,Pilati2008a} and the analysis of an experiment in Ref.~\cite{Shin2008a},
although the value differs somewhat from the predictions in Refs.~\cite{Combescot2007a,Combescot2008a}
and observations in new experiments~\cite{Nascimbene2009a,Nascimbene2010a}.
Neglecting the imaginary parts of the self-energy gives the zero-momentum polaron energy
$-0.70\,E_\mathrm{F}$ and the effective mass $m^* \approx 1.47\,m$.

\section{Contact}
\label{sec:contact}


The self-energy yields also the momentum distribution through the 
relation~\cite{Combescot2009a}
\begin{equation}
  n_{{\bf k}} = \int \frac{dk_0}{2\pi} G({\bf k},k_0),
\end{equation}
but for this one would need to determine the Green's function and hence
also the self-energy for all frequencies $k_0$. However, I propose
that for a dilute gas the high momentum asymptote of the distribution can be
calculated using a simple perturbation theory.

Consider an adiabatic switching on of the interactions,
$H(t) = H_0 + Ve^{-\eta t}$ for $t < 0$ and $\eta \rightarrow 0$, 
starting from the ground state of $H_0$. According to Gell-Mann and Low
this yields the ground state of $H$ as
\begin{equation}
    |\psi\rangle = S(0,-\infty) |\psi_0\rangle,
\end{equation}
where $|\psi_0\rangle$ is the ground state of $H_0$ and the S-matrix 
\begin{equation}
   S(t,t') = \mathcal{T} e^{-i\int_{t'}^t dt_1 \, V(t_1)},
\end{equation}
where $V(t_1)$ is the interaction part of the total Hamiltonian
expressed in the interaction picture.

The expectation value of the occupation number for the spin state $\uparrow$
in some momentum state $n_{{\bf q},\uparrow} = \langle \psi| c_{{\bf q},\uparrow}^\dagger c_{{\bf q},\uparrow} |\psi \rangle$ in 
the interacting ground state is now
\begin{equation}
   n_{{\bf q},\uparrow} = \left| c_{{\bf q},\uparrow} S(0,-\infty) |\psi_0\rangle \right|^2.
\end{equation}
Motivated by the arguments of Shina Tan~\cite{Tan2008a,Tan2008b}, I assume that only pair-wise 
correlations are relevant in the high momentum limit $q \gg k_\mathrm{F}$, 
and thus the observation of a particle at the momentum ${\bf q}$ implies that 
some pair of atoms (with momenta ${\bf p}$ and ${\bf k}$) in the initial 
state $|\psi_0\rangle$ have
scattered to the states ${\bf q}$ and ${\bf p+k-q}$. The probability amplitude 
for such scattering is provided by the many-body scattering T-matrix 
$\Gamma (P)$ (which, according to Section~\ref{sec:scattering} depends only 
on the center-of-mass four-momentum $P$) divided by the energy change due 
to the scattering, i.e. the probability is
\begin{equation}
   P_\uparrow ({\bf k},{\bf p};{\bf q}) = \left| \frac{\Gamma (P)}{\varepsilon_{\bf p+k-q}^* + \varepsilon_q^* - \varepsilon_p - \varepsilon_k}\right|^2 n_k n_p  \approx  \left| \frac{\Gamma (P)}{2\epsilon_q}\right|^2 n_k n_p,
\end{equation}
where the occupation numbers $n_k$ and $n_p$ give the probability that the momentum states ${\bf k}$ and ${\bf p}$ were initially
occupied and I have assumed that $q \gg p,k$ and that $\varepsilon_q \approx \epsilon_q$ (based on the observation that the Hartree 
energy shift vanishes for large momenta $q$). This is similar to the 
first-order perturbation theory result for the transition 
probability assuming the coupling strength $\Gamma(P)$ and the 
energy difference $2\epsilon_q$. However, notice that since 
the coupling strength is in reality a T-matrix, it includes 
scatterings to arbitrary order but the correlations are limited
to the two-particle level.

However, the observation of a particle at the momentum ${\bf q}$ tells
only that some atoms have scattered but does not specify the original 
momenta of the atoms ${\bf p, k}$.  
On the other hand, since higher order correlations are ignored, there 
are holes left at the original momenta ${\bf p, k}$. 
Another trajectory leading into observation
of a particle at momentum ${\bf q}$ would be one in which two atoms from 
some other momentum states ${\bf p',k'}$ scatter to states 
${\bf q,p'+k'-q}$. However,
the two trajectories yield a zero overlap when calculating the expectation 
value since the final states are mutually orthogonal due to the different
configuration of holes. Hence the two trajectories do not interfere 
coherently and one expects that the different trajectories need to be 
summed incoherently when calculating the expectation value of the observable $c_{\bf q}^\dagger c_{\bf q}$. Therefore I assume that the occupation number at 
some high momentum state $q \gg k_\mathrm{F}$ is given by
\begin{equation}
   n_{{\bf q},\uparrow}  \approx \frac{1}{V^2} \sum_{\bf k,p} \left| P({\bf k},{\bf p}; {\bf q})\right| \approx \frac{\sum_{\bf k,p} \left|\Gamma (P)\right|^2 n_k n_p}{4\epsilon_q^2},
\end{equation}

The contact~\cite{Tan2008a,Tan2008b}, defined as the limit $C = \lim_{q \rightarrow \infty} q^4 n_{\bf q}$, is now
\begin{equation}
   C = \left(\frac{m}{\hbar^2}\right)^2 \frac{1}{V^2}\sum_{{\bf k},{\bf p}} |\Gamma(P)|^2 n_{\uparrow,k} n_{\downarrow,p}.
\label{eq:contact}
\end{equation}
The value of the contact depends strongly on the low momentum occupation 
numbers $n_k$ both directly but also indirectly through the many-body 
scattering T-matrix $\Gamma$. Notice that this equation differs from the one 
used for example in Refs.~\cite{Sartor1980a,Combescot2009a}
where the occupation numbers were defined through self-energies. It 
is instructive to consider the weakly interacting limit first. Approximating
$\Gamma \approx \frac{4\pi\hbar^2 a}{m}$ yields for the contact
\begin{eqnarray}
   C \approx \left(\frac{m}{\hbar^2}\right)^2 \frac{1}{V^2} \left(\frac{4\pi\hbar^2 a}{m}\right)^2 \sum_{{\bf k},{\bf p}} n_{\uparrow,k} n_{\downarrow,p} = 4\pi^2 \left(na\right)^2,
\end{eqnarray}
where $n$ is the total atom density of a balanced gas. This is the exact
result for a weakly interacting gas~\cite{Sartor1980a,Tan2008b}. 

Another simple system for which one can easily determine the contact is the
polaron system where the majority component atoms form an ideal Fermi sphere 
$n_{\uparrow,k} = \theta(k_\mathrm{F}-k)$ and the minority component atoms
consist only of a single impurity at the zero momentum $n_{\downarrow,p} = \delta_{p0}$.
The contact becomes now
\begin{equation}
   C = \left(\frac{m}{\hbar^2}\right)^2 \frac{1}{V^2} \sum_{\bf k} |\Gamma(\frac{\bf k}{2},\frac{\varepsilon_{\downarrow,0} + \varepsilon_{\uparrow,k}}{2})|^2 n_{\uparrow,k}.
\end{equation}
Notice that this quantity depends on the Hartree shifts at all momenta because
these affect the intermediate states in the many-body scattering T-matrix.
At unitarity I obtain $C = 0.10\,k_{\uparrow,\mathrm{F}} k_{\downarrow,\mathrm{F}}^3$
(or $C = 0.13 \, k_{\uparrow,\mathrm{F}} k_{\downarrow,\mathrm{F}}^3$ when neglecting
the imaginary part in the self-energy). This
is slightly higher than the value $C \approx 0.08\,k_{\uparrow,\mathrm{F}} k_{\downarrow,\mathrm{F}}^3$ 
obtained in~\cite{Punk2009a}. I am not aware
of experimental results for the polaron contact.

The contact for a balanced system has been determined experimentally 
using various methods but the problem with the present theory is that 
it does not give access to occupation numbers $n_k$ for low momenta $k \sim k_{\mathrm{F}}$.
The simplest assumption is that both atomic components form an ideal Fermi sphere with
occupation numbers $n_k = \theta(k_\mathrm{F}-k)$. Such assumption gives for the value
of the contact $C \approx 0.064\,k_\mathrm{F}^4$. Omitting the imaginary parts in the
self-energy yields $C \approx 0.14\,k_\mathrm{F}^4$. 
Obviously, any finite value for the
contact is incompatible with the assumption of a Fermi-Dirac-like distribution
for the occupation numbers as it has an exponentially decaying tail in violation 
with the Bethe-Peierls condition. 
The assumption can be easily improved by making an ansatz for the occupation numbers using
a functional form that has the correct asymptote. For this end 
I use the BPC ansatz
\begin{equation}
   n_q = \frac{1}{2} \left( 1-\frac{\epsilon_q-\mu}{\sqrt{(\epsilon_q-\mu)^2 + \Delta^2}} \right),
\label{eq:bcsoccu}
\end{equation}
where $\mu$ and $\Delta$ are considered \emph{fitting parameters} with $\mu$ chosen
so that the atom numbers are fixed and $\Delta$ is obtained from the value of
the contact $C = \lim_{k\rightarrow \infty} q^4 n_q = \left(\frac{m}{\hbar^2}\right)^2 \Delta^2$.
This requires an iterative solution so that the contact obtained
from Eq.~\eqref{eq:contact} matches the contact from Eq.~\eqref{eq:bcsoccu}.
Using this ansatz I obtain $\Delta \approx 0.49\,E_\mathrm{F}$, $\mu \approx 0.79\,E_\mathrm{F}$ and $C \approx 0.060\,k_\mathrm{F}^4$.
In addition, the Hartree energy shift at the Fermi momentum equals $-0.34\,E_\mathrm{F}$.
Neglecting the imaginary parts of the self-energy yields $\Delta \approx 0.64\,E_\mathrm{F}$, $\mu \approx 0.66\,E_\mathrm{F}$, 
$C \approx 0.103\,k_\mathrm{F}^4$, and the Hartree shift at the Fermi momentum $-0.46\,E_\mathrm{F}$.

The difference in the values from the two approaches for the self-consistency (as discussed in Section~\ref{sec:hartree})
can be understood in terms of the hole lifetime. Including the imaginary parts of the self-energy in the
Green's functions of the scattering particles $p$ and $k$ will decrease the expectation value of the 
scattering probability since the system cannot return to the initial state if the holes have decayed.
In the extreme limit of a very short lifetime, or very large imaginary parts of the self-energy, all
scatterings will be blocked analogously to the quantum Zeno effect. 

Considering the simplicity of the present method, these figures are in surprisingly good agreement 
with Monte Carlo results~\cite{Gandolfi2010a} (yielding $C=0.1147(3)k_\mathrm{F}^4$), self-consistent 
Green's function method~\cite{Haussmann2009a} (yielding the pairing gap $\Delta=0.46\,E_\mathrm{F}$, the Hartree
shift $U = -0.50\,E_\mathrm{F}$, and the contact $C = 0.102\,k_\mathrm{F}^4$), and
experiments (Ref.~\cite{Navon2010a} yielding $C \approx 0.12\,k_\mathrm{F}^4$ and 
Ref.~\cite{Kuhnle2010a} yielding $C \approx 0.096\,k_\mathrm{F}^4$).

\section{Discussion on the model}
\label{sec:discussion}

The key approximations in this work are the Green's function ansatz in Eq.~\eqref{eq:green} and the
neglect of the pairing part of the self-energy as described in Section~\ref{sec:hartree_initial}.
The justification of both of these approximations is based on the argument that they are sufficient
for describing the Hartree energy shift but one should not make too far reaching conclusions based
on this model. However, as seen by comparing the results obtained for the zero-temperature Fermi sphere ansatz
and for the BPC ansatz in Figs.~\ref{fig:hartree_unitary} and~\ref{fig:hartree_unitary_noimag}, and
also the values obtained for the contact in Section~\ref{sec:contact}, the values of the
Hartree self-energy do not have a very dramatic dependence on the ansatz. A distinct Fermi 
surface does make features sharper, see Fig.~\ref{fig:hartree_unitary_noimag}, but qualitatively
the behavior is similar and the actual numerical values close to the Fermi momentum are 
in the same range. Thus one would expect that changing the actual form of the BPC ansatz would not
change the values much. One interesting possibility would be an ansatz that combines a step-like
behavior at the Fermi surface (albeit with a step smaller than unity) and a proper $1/k^4$ high
momentum asymptote or, even better, describing the low momentum regime with the finite temperature
Fermi-Dirac distribution and continuously merging this with a $1/k^4$ tail. The result of
such an ansatz would be a Hartree shift lying somewhere between the two ansatzes used in this work.

The BPC ansatz and the related gapless excitation spectrum implied by Eq.~\eqref{eq:green}
is in contradiction with the expected pseudogap physics at high temperatures~\cite{Chen2005a}
but this is a direct consequence of the choice of the ansatz. However, despite this deficit,
the ansatz used here is in agreement with experiments that have been used for trying to observe 
the pseudogap phenomena~\cite{Gaebler2010a,Leskinen2010a}.

The other key approximation, namely neglecting the pairing part, is more critical in some
cases than in others. Considering the values of the contact calculated in Section~\ref{sec:contact} 
as a benchmark, it appears that the full model in which also the imaginary part of the Hartree
self-energy is included in the iteration works better in the case of the polaron than in the
balanced system, whereas omitting the imaginary part works better in the balanced case than 
in the polaron case. This behavior is not at all surprising, considering that in the case of
polarons, the many-body pair formation is strongly suppressed by the huge mismatch in the Fermi
momenta. Hence the pairing part can be safely ignored and the Hartree self-energy contains
all the relevant physics, including lifetime effects. Notice that the same argument holds also 
at sufficiently high temperatures when the pair formation is suppressed by thermal excitations,
even in the balanced gas.
In the case of a balanced gas at low temperatures, the pairing
part of the self-energy will also yield an imaginary part that will at least partly cancel the
imaginary parts of the Hartree energy shift. In particular, close to the Fermi surface the
cancellation is expected to be almost complete~\cite{Luttinger1961a}. Thus neglecting the
imaginary part of the Hartree self-energy in the case of a balanced gas is probably well
justified.

\section{Conclusions}
\label{sec:summary}

In this manuscript I have calculated the Hartree self-energy of strongly 
interacting fermionic atoms. Short range of the atom-atom interaction 
potential and the diluteness of the
gas allows an efficient description of the interactions through the 
momentum dependent scattering amplitude.
I have calculated the Hartree energy shifts, effective mass, and the 
value of the contact for both balanced and spin-imbalanced gases.
For the polaron system these values are in an excellent
agreement with the results from other theories, including Monte Carlo results, and with experimental
results, although it is not clear what value of the effective mass should be used as a reference. 
For the balanced system the agreement with more exact methods is also good when 
using Green's function ansatzes that satisfy the Bethe-Peierls condition. In the future, it will 
be interesting to consider the connection
of the present theory with a BCS-type pairing theory by including also the pair part of the 
self-energy $\Sigma$. Furthermore, the generalization of the theory to inhomogeneous systems will be
useful, allowing the treatment of the Hartree energy shift properly in Bogoliubov-deGennes-type 
spatially dependent mean-field theories~\cite{Ohashi2005a,Jensen2007a}.

{\ack
I would like to acknowledge enlightening discussions with M.O.J. Heikkinen, J. Kajala, and D.-H. Kim.}

\appendix

\section{Derivation of the many-body scattering T-matrix}
\label{app:scatamp}

Here I will review the standard derivation of the many-body scattering T-matrix
and how it can be expressed through the two-particle scattering amplitude.
The calculation follows closely the derivation in Ref.~\cite{FetterAndWalecka}
but here the discussion is kept more general as all the key approximations will
be done in the main text.

In the ladder approximation the many-body scattering T-matrix can be written as
\begin{eqnarray}
\fl  \Gamma(P_1,P_2;P_3,P_4) = &U_0(P_1-P_3) + i \int \frac{dQ}{(2\pi)^4}\, U_0(Q) \nonumber \\
  &\times G(P_1-Q) G(P_2+Q) \Gamma(P_1-Q,P_2+Q;P_3,P_4),
\label{eq:gamma1}
\end{eqnarray}
where $U_0(P)$ is the interaction potential. Here $P_1,P_2$ are the incoming
four-momenta of the two particles and $P_3,P_4$ are the corresponding outgoing momenta
(and while the above equation has been written without spin-indices, it is to be
understood that $P_1$ and $P_3$ describe incoming and outgoing four-momenta
of one spin state and $P_2$ and $P_4$ describe the other spin state).
This form for the many-body scattering T-matrix is not particularly useful since
the potential $U_0(P)$ is seldom known. However, for short range interaction potentials the 
T-matrix obtains a simple form when expressed using the scattering amplitude.

\subsection{Scattering amplitude}

I will first review the standard scattering calculation of two particles of 
mass $m$ and (relative) incoming momentum ${\bf k}$ scattering from a spherically 
symmetric potential $U_0(r)$. I will restrict the analysis to the $s$-wave 
scattering, which is the relevant limit for low energy scatterings.

The scattering of a particle, with incoming momentum ${\bf k}$, from a spherically 
symmetric potential $U_0({\bf x})$ is described by the Schr\"odinger equation
\begin{equation}
    \frac{\hbar^2}{2m_\mathrm{red}} \left( \nabla^2 + k^2 \right) \psi_{\bf k}({\bf x})= \frac{\hbar^2}{m} \left( \nabla^2 + k^2 \right) \psi_{\bf k}({\bf x}) = U_0({\bf x}) \psi_{\bf k}({\bf x}),
\end{equation}
where ${\bf x}$ is the distance between the particles and the reduced
mass $m_\mathrm{red} = m/2$ is used in order to be able to describe the scattering 
of two particles with equal masses $m$. This equation can be rewritten as an 
integral equation
\begin{equation}
    \psi_{\bf k} ({\bf x}) = e^{i {\bf k} \cdot {\bf x}} - \int d{\bf y}\, \frac{m}{4\pi\hbar^2} \frac{e^{ik|{\bf x} - {\bf y}|}}{|{\bf x} - {\bf y}|} U_0({\bf y}) \psi_{\bf k} ({\bf y}).
\end{equation}
This can be expressed in the momentum space by taking the Fourier transform
\begin{equation}
\fl    \psi_{\bf k} ({\bf p}) = (2\pi)^3 \delta({\bf p} - {\bf k}) - \frac{1}{\frac{\hbar^2}{m} p^2 - \frac{\hbar^2}{m} k^2 - i\eta} \int \frac{\bf dq}{(2\pi)^3} \, U_0({\bf q}) \psi_{\bf k}({\bf p} - {\bf q}).
\label{eq:scatwf}
\end{equation}

Defining the scattering amplitude
\begin{equation}
-\frac{4\pi \hbar^2}{m} f({\bf s},{\bf k}) = \int \frac{d{\bf q}}{(2\pi)^3}\, U_0({\bf q}) \psi_{\bf k} ({\bf s}-{\bf q}),
\label{eq:scatamp0}
\end{equation}
the scattering wavefunction becomes 
\begin{equation}
    \psi_{\bf k} ({\bf p}) = (2\pi)^3 \delta({\bf p} - {\bf k}) + \frac{\frac{4\pi\hbar^2}{m}f({\bf p},{\bf k})}{2\epsilon_p - 2\epsilon_k - i\eta},
\label{eq:scatwf2}
\end{equation}
where $\epsilon_p = \frac{\hbar^2}{2m} p^2$.
The key quantity here is the scattering amplitude $f({\bf k'},{\bf k})$ and
the goal is to formulate the many-body scattering T-matrix using the scattering amplitude.
However, the scattering T-matrix must yield the above scattering wavefunction
for elastic scatterings and in the absence of many-body contributions.

For future purposes, it is worthwhile to write down the completeness relation,
which expresses the completeness of the scattering states in the absence of
bound states, namely
\begin{equation}
\int \frac{d{\bf k}}{(2\pi)^3}\, \psi_{\bf k} ({\bf p}) \psi_{\bf k} ({\bf p'})^* = (2\pi)^3 \delta ({\bf p} - {\bf p'}).
\label{eq:complete}
\end{equation}
The completeness relation, together with Eq.~\eqref{eq:scatamp0} yields
\begin{equation}
  -\int \frac{d{\bf k}}{(2\pi)^3} \, \frac{4\pi\hbar^2}{m} f({\bf p,k}) \psi_{\bf k} ({\bf p'}) = U_0({\bf p-p'}).
\end{equation}
Combined with Eq.~\eqref{eq:scatwf2} this yields for a hermitian potential $U_0$
\begin{equation}
\fl  f({\bf p,p'}) - f({\bf p',p})^* = -\int \frac{d{\bf k}}{(2\pi)^3} \, \frac{4\pi\hbar^2}{m} f({\bf p,k})  f({\bf p',k})^* \left( \frac{1}{2\epsilon_k - 2\epsilon_p + i\eta} - \frac{1}{2\epsilon_k - 2\epsilon_{p'} - i\eta} \right),
\label{eq:opticaltheory}
\end{equation}
which can be considered as a generalization of the optical theorem for the inelastic scattering amplitude.

For short range potentials and elastic scatterings ($k = s$), the scattering amplitude is 
well approximated by 
\begin{equation}
  f({\bf s},{\bf k}) \approx \frac{-a}{1+ika+\frac{1}{2}r_\mathrm{eff} ak^2}, \quad \mathrm{for}\, k=s
\end{equation}
where $r_\mathrm{eff}$ is called the effective range and it is of the same order
of magnitude as the interaction potential range. For a dilute atomic gas in the vicinity
of a broad Feshbach resonance the factor $kr_\mathrm{eff} \ll 1$ and the effective
range correction can be neglected. Indeed, for a \emph{zero-range} delta function 
pseudopotential interaction
$U_0(r) = \frac{4\pi\hbar^2 a}{m} \delta (r) \frac{d}{dr} \left(r \cdot \right)$ the
scattering amplitude becomes
\begin{equation}
  f({\bf s},{\bf k}) \approx \frac{-a}{1+ika}, \quad \mathrm{for}\, k=s.
\label{eq:scatamp}
\end{equation}
However, while the scattering calculation is usually restricted to determining the
elastic scattering amplitude, the inelastic scattering amplitude will be needed in the
main text. I am not aware of a result for the inelastic scattering amplitude for the
above delta function pseudopotential interaction, but I assume that the inelastic 
scattering amplitude depends only on the magnitude of the incoming relative momentum, i.e.
\begin{equation}
  f({\bf s},{\bf k}) = f({\bf k}) \approx \frac{-a}{1+ika}, \quad \mathrm{for\,all}\,{\bf s,k}.
\label{eq:scatampinel}
\end{equation}
Such assumption for the inelastic scattering amplitude is consistent
at unitarity $a \rightarrow \infty$ in the sense that it satisfies Eq.~\eqref{eq:opticaltheory}, 
but it is not necessarily true for any $a$ and certainly not true for arbitrary potentials.

\subsection{Many-body scattering T-matrix}

The T-matrix in Eq.~\eqref{eq:gamma1} can be expressed using a two-particle wavefunction $\mathcal{Q}$ as
\begin{equation}
    \Gamma (P_1,P_2;P_3,P_4) = \int \frac{dQ}{(2\pi)^4}\, U_0(Q) \mathcal{Q}(P_1-Q,P_2+Q;P_3,P_4),
\end{equation}
which satisfies the ladder approximation for $\Gamma$ in Eq.~\eqref{eq:gamma1}
if $\mathcal{Q}$ obeys
\begin{eqnarray}
   \mathcal{Q}(P_1,P_2;P_3,P_4) &= (2\pi)^4 \delta(P_1-P_3) + i G(P_1) G(P_2) \nonumber \\
&\times \int \frac{dQ}{(2\pi)^4}\, U_0(Q) \mathcal{Q} (P_1-Q,P_2+Q;P_3,P_4),
\label{eq:Qfunc1}
\end{eqnarray}
which can be easily verified by substitution. The two-particle wavefunction
is used here only in an auxiliary role and will soon be replaced by the pair
susceptibility $\chi$.

The two-body scattering processes are often easiest to describe using the center-of-mass 
coordinates. The total center-of-mass four-momentum satisfies
\begin{equation}
  2P = P_1 + P_2 = P_3 + P_4,
\end{equation}
where the latter equality follows from the conservation of the total four-momentum 
in a homogeneous system (homogeneous also in time, hence the conservation
of the frequency). The relative four-vectors are
$\delta P = \frac{1}{2} (P_1-P_2) =: S$ and $\delta P' = \frac{1}{2} (P_3-P_4) =: K$.
The two-particle wavefunction expressed using these variables is now
\begin{eqnarray}
   \fl \mathcal{Q}(&P + S, P - S;P + K, P - K) = (2\pi)^4 \delta(S - K) + i G(P + S) \nonumber \\
\fl &\times G(P - S) \int \frac{dQ}{(2\pi)^4}\, U_0(Q) \mathcal{Q}(P + S -Q,P - S +Q;P + K,P - K)
\end{eqnarray}
Assuming that the interaction potential $U_0(Q)$ does not depend on the 
frequency (time-independent interaction potential), the frequency integral 
can be calculated in the right hand side, yielding
\begin{eqnarray}
  \fl \mathcal{Q}(P + S, P - S;&P + K, P - K) = (2\pi)^4 \delta(S - K) \nonumber \\
\fl &+ i G(P + S) G(P - S) \int \frac{d{\bf q}}{(2\pi)^3}\, U_0({\bf q}) \chi(S - {\bf q},K;P).
\label{eq:Qfunc2}
\end{eqnarray}
where the pair susceptibility $\chi$ is defined as
\begin{eqnarray}
  \chi(S - {\bf q}, K; P) &= \int \frac{dq_0}{2\pi} \,\mathcal{Q}(P + S-Q,P - S+Q;P + K,P - K) \nonumber \\
&= \chi ({\bf s} - {\bf q}, K; P).
\end{eqnarray}
The last equality follows from the observation that the pair susceptibility $\chi$
does not depend on the frequency of the relative wave vector $S$.
This is obvious if one does the change of integration variable 
$s_0 - q_0 = x$. The equation does not depend on $k_0$ either, as will become
apparent soon.

Substituting Eq.~\eqref{eq:Qfunc2} into $\chi$ yields
\begin{eqnarray}
 \chi({\bf s}, K; P) = (2\pi)^3 \delta ({\bf s}- {\bf k}) &+ i \int \frac{dq_0}{2\pi} \,G(P+{\bf s} - q_0) G(P - {\bf s} + q_0) \nonumber \\
&\times \int \frac{d{\bf q}}{(2\pi)^3} \, U_0({\bf q}) \chi ({\bf s} - {\bf q}, K; P).
\end{eqnarray}
This equation has no dependence on $k_0$: it is the defining equation
for the function $\chi$, and the function definition remains exactly the same
no matter what value is used for $k_0$. Choosing $k_0  = 0$ yields
\begin{eqnarray}
\fl  \chi({\bf s}, {\bf k}; P) = (2\pi)^3 \delta ({\bf s}- {\bf k}) + I({\bf s};P) \int \frac{d{\bf q}}{(2\pi)^3} \, U_0({\bf q}) \chi ({\bf s} - {\bf q}, {\bf k}; P).
\end{eqnarray}
Not wanting to do key approximations in the Appendix, I have denoted here
\begin{equation}
  I({\bf s};P) = i \int \frac{dq_0}{2\pi} \,G(P+{\bf s} - q_0) G(P - {\bf s} + q_0).
\end{equation}

The scattering T-matrix can now be written in terms of $\chi$ as
\begin{eqnarray}
\fl    \Gamma (P_1,P_2;P_3,P_4) &= \int \frac{d{\bf q}}{(2\pi)^3}\, U_0({\bf q}) \int \frac{dq_0}{2\pi} \, \mathcal{Q}(P_1-Q,P_2+Q;P_3,P_4) \nonumber \\
&= \int \frac{d{\bf q}}{(2\pi)^3}\, U_0({\bf q}) \int \frac{dq_0}{2\pi} \, \mathcal{Q}(P + S - Q,P - S +Q;P + K,P - K) \nonumber \\
&= \int \frac{d{\bf q}}{(2\pi)^3}\, U_0({\bf q}) \chi ({\bf s} - {\bf q}, {\bf k}; P) \nonumber \\
&= \Gamma ({\bf s}, {\bf k}; P).
\label{eq:gammavschi}
\end{eqnarray}
Notice that the scattering T-matrix does not depend on the relative 
frequencies $s_0$ and $k_0$ because the pair susceptibility
has no such dependence.

\subsection{Two-body scattering T-matrix}

In order to obtain the connection to the scattering amplitude, one needs to first take
the explicit two-particle limit of the many-body scattering T-matrix. In the case of
the elastic scattering, this will yield the scattering amplitude.
The two-body scattering T-matrix $\Gamma_0$ is defined as the 
limit of $\Gamma$ in which all the Green's functions $G(K)$ are replaced by the non-interacting
Green's functions $G_0(K) = 1/(k_0 - \epsilon_k)$ with the free particle dispersion
$\epsilon_k = \hbar^2 k^2/2m$. The two-body scattering T-matrix is thus defined as
\begin{equation}
   \Gamma_0({\bf s}, {\bf k}; P) = \int \frac{d{\bf q}}{(2\pi)^3}\, U_0({\bf q}) \chi_0 ({\bf s} - {\bf q}, {\bf k}; P),
\end{equation}
where 
\begin{eqnarray}
\fl  \chi_0({\bf s}, {\bf k}; P) = (2\pi)^3 \delta ({\bf s}- {\bf k}) - \frac{1}{E_P - 2\epsilon_{s}} \int \frac{d{\bf q}}{(2\pi)^3} \, U_0({\bf q}) \chi_0 ({\bf s} - {\bf q}, {\bf k}; P),
\label{eq:chi0def}
\end{eqnarray}
and $E_P = 2p_0 - \frac{\hbar^2 {\bf p}^2}{m}$. 
Multiplying $\chi_0$ by $(E_P - 2\epsilon_{s})$ gives
\begin{eqnarray}
\fl  (E_P - 2\epsilon_{s}) &\chi_0({\bf s}, {\bf k}; P) - \int \frac{d{\bf q}}{(2\pi)^3} \, U_0({\bf q}) \chi_0 ({\bf s} - {\bf q}, {\bf k}; P) = \nonumber \\
\fl &(E_P - 2\epsilon_{s}) (2\pi)^3 \delta ({\bf s}- {\bf k}) = (E_P - 2\epsilon_{k}) (2\pi)^3 \delta ({\bf s}- {\bf k}).
\label{eq:chi0}
\end{eqnarray}
In order to obtain the connection to the scattering amplitude, the $\chi_0$ can be
expressed as
\begin{equation}
   \chi_0({\bf s}, {\bf k};P) = (E_P - 2\epsilon_{k}) \int \frac{d{\bf \tilde k}}{(2\pi)^3}\, \frac{\psi_{\bf \tilde k} ({\bf s}) \psi_{\bf \tilde k} ({\bf k})^*}{E_P - 2\epsilon_{\tilde k}}.
\label{eq:chi1}
\end{equation}
This can be confirmed by substituting this into Eq.~\eqref{eq:chi0} and remembering two important equations: the Schr\"odinger equation
in the momentum space (Eq.~\eqref{eq:scatwf})
\begin{equation}
   \psi_{\bf k} ({\bf p}) = (2\pi)^3 \delta ({\bf p} - {\bf k}) - \frac{1}{2\epsilon_p - 2\epsilon_k - i\eta} \int \frac{d{\bf q}}{(2\pi)^3} \, U_0({\bf q}) \psi_{\bf k} ({\bf p} - {\bf q}),
\label{eq:schrode}
\end{equation}
and the completeness relation (Eq.~\eqref{eq:complete})
\begin{equation}
   \int \frac{d{\bf k}}{(2\pi)^3} \, \psi_{\bf k} ({\bf p}) \psi_{\bf k} ({\bf p'})^* = (2\pi)^3 \delta ({\bf p} - {\bf p'}).
\end{equation}

Substituting Eq.~\eqref{eq:chi1} into Eq.~\eqref{eq:chi0} yields
\begin{eqnarray}
\fl(2\pi^3) \delta ({\bf s} - {\bf k}) = & (E_P - 2\epsilon_{s}) \int \frac{d{\bf \tilde k}}{(2\pi)^3} \frac{\psi_{\bf \tilde k} ({\bf s}) \psi_{\bf \tilde k} ({\bf k})^*}{E_P - 2\epsilon_{\tilde k}} \nonumber \\
&- \int \frac{d{\bf q}}{(2\pi)^3} U_0({\bf q}) \int \frac{d{\bf \tilde k}}{(2\pi)^3} \frac{\psi_{\bf \tilde k} ({\bf s} - {\bf q}) \psi_{\bf \tilde k} ({\bf k})^*}{E_P - 2\epsilon_{\tilde k}} \nonumber \\
= & \int \frac{d{\bf \tilde k}}{(2\pi)^3} \left[ (E_P - 2\epsilon_{s}) \frac{\psi_{\bf \tilde k} ({\bf s}) \psi_{\bf \tilde k} ({\bf k})^*}{E_P - 2\epsilon_{\tilde k}} \right. \nonumber \\
&\left.- \left( \int \frac{d{\bf q}}{(2\pi)^3} U_0({\bf q}) \psi_{\bf \tilde k} ({\bf s} - {\bf q}) \right) \frac{\psi_{\bf \tilde k} ({\bf k})^*}{E_P - 2\epsilon_{\tilde k}} \right] \nonumber \\
=&
\int \frac{d{\bf \tilde k}}{(2\pi)^3} \left[ (E_P - 2\epsilon_{s}) \frac{\psi_{\bf \tilde k} ({\bf s}) \psi_{\bf \tilde k} ({\bf k})^*}{E_P - 2\epsilon_{\tilde k}} \right. \nonumber \\
&\left.- (2\epsilon_{\tilde k} - 2\epsilon_{s}) \psi_{\bf \tilde k} ({\bf p}) \frac{\psi_{\bf \tilde k} ({\bf k})^*}{E_P - 2\epsilon_{\tilde k}} \right] \nonumber \\
= & \int \frac{d{\bf \tilde k}}{(2\pi)^3} \frac{\psi_{\bf \tilde k} ({\bf s}) \psi_{\bf \tilde k} ({\bf k})^*}{E_P - 2\epsilon_{\tilde k}} \left[ (E_P - 2\epsilon_{s}) - (2\epsilon_{\tilde k} - 2\epsilon_{s}) \right] \nonumber \\
= & \int \frac{d{\bf \tilde k}}{(2\pi)^3} \psi_{\bf \tilde k} ({\bf s}) \psi_{\bf \tilde k} ({\bf k})^* \nonumber \\
=&
(2\pi)^3 \delta ({\bf s} - {\bf k}).
\end{eqnarray}

Eq.~\eqref{eq:chi1} provides now the required connection between the T-matrix formulation
and the scattering calculation in~\ref{app:scatamp}. There the
solution of the Schr\"odinger equation Eq.~\eqref{eq:scatwf} was connected
to the scattering amplitude by Eq.~\eqref{eq:scatwf2}
\begin{equation}
   \psi_{\bf k} ({\bf p}) = (2\pi)^3 \delta ({\bf p} - {\bf k}) + \frac{\frac{4\pi \hbar^2}{m} f ({\bf p}, {\bf k})}{2\epsilon_k - 2\epsilon_p + i \eta},
\end{equation}
yielding now
\begin{equation}
\fl   \chi_0({\bf s}, {\bf k};P) = \psi_{{\bf k}} ({\bf s}) + \frac{4\pi\hbar^2}{m} \int \frac{d{\bf \tilde k}}{(2\pi)^3} \psi_{\bf \tilde k} ({\bf s}) \left( \frac{1}{E_P - 2\epsilon_{\tilde k}} + \frac{1}{2\epsilon_{\tilde k} - 2\epsilon_{k} - i\eta} \right) f ({\bf k}, {\bf \tilde k})^*.
\label{eq:chi2}
\end{equation}

Similarly the two-body scattering T-matrix can now be expressed using the scattering
amplitude as
\begin{eqnarray}
\fl   \Gamma_0(&{\bf s}, {\bf k}; P) = \int \frac{d{\bf q}}{(2\pi)^3}\, U_0({\bf q}) \psi_{{\bf k}} ({\bf s} - {\bf q}) \\
\fl & - \frac{4\pi\hbar^2}{m} \int \frac{d{\bf q}}{(2\pi)^3}\, U_0({\bf q}) \int \frac{d{\bf \tilde k}}{(2\pi)^3} \psi_{\bf \tilde k} ({\bf s} - {\bf q}) \left( \frac{1}{E_P - 2\epsilon_{\tilde k}} + \frac{1}{2\epsilon_{\tilde k} - 2\epsilon_{k} - i\eta} \right) f ({\bf k}, {\bf \tilde k})^* \nonumber \\
\fl &= -\frac{4\pi \hbar^2}{m} f({\bf s}, {\bf k}) +
\left( \frac{4\pi \hbar^2}{m} \right)^2 \int \frac{d{\bf \tilde k}}{(2\pi)^3} f({\bf s}, {\bf \tilde k}) \left( \frac{1}{E_P - 2\epsilon_{\tilde k}} + \frac{1}{2\epsilon_{\tilde k} - 2\epsilon_{k} - i\eta} \right) f ({\bf k}, {\bf \tilde k})^*.\nonumber 
\label{eq:gamma0vsscatamp}
\end{eqnarray}
A familiar form can be obtained when considering scatterings on-the-energy shell 
(scatterings that conserve energy). These correspond to the limit 
$E_P = 2\epsilon_{k} + i\eta$, for which one obtains
\begin{equation}
   \Gamma_0 ({\bf s}, {\bf k};P) = -\frac{4\pi \hbar^2}{m} f({\bf s}, {\bf k}).
\end{equation}

\subsection{Many-body scattering T-matrix revisited}

Here I will show the explicit connection between the many-body scattering T-matrix $\Gamma$ 
and the two-body scattering T-matrix $\Gamma_0$, and hence also between $\Gamma$ and the 
scattering amplitude $f$. This is of course important since the self-energy $\Sigma$ 
is defined through $\Gamma$.

The full many-body scattering $T$-matrix $\Gamma$ was expressed using
the pair susceptibility $\chi$ through Eq.~\eqref{eq:gammavschi}.
The $\chi$ can be rewritten as
\begin{eqnarray}
\fl  \chi({\bf s}, {\bf k}; P) - \frac{1}{E_P-2\epsilon_s} \int &\frac{d{\bf q}}{(2\pi)^3} \, U_0({\bf q}) \chi ({\bf s}-{\bf q},{\bf k};P) \\
\fl &= (2\pi)^3 \delta ({\bf s}- {\bf k}) \nonumber + \left[ I({\bf s};P) - \frac{1}{E_P-2\epsilon_s}\right] \Gamma ({\bf s}, {\bf k};P) \nonumber \\
\fl &= (2\pi)^3 \delta ({\bf s}- {\bf k}) + \tilde \chi({\bf s};P) \Gamma ({\bf s},{\bf k};P),
\label{eq:fullchi}
\end{eqnarray}
where the notation has been simplified by defining the regularized pair susceptibility
\begin{equation}
  \tilde \chi ({\bf s};P) = I({\bf s};P) - \frac{1}{E_P-2\epsilon_s}.
\label{eq:tildechi}
\end{equation}
Recalling the definition of $\chi_0$ (Eq.~\eqref{eq:chi0def}), the operator
part on the left hand side of Eq.~\eqref{eq:fullchi} is equal to the inverse of $\chi_0$,
yielding the formal operator equation
\begin{eqnarray}
   \chi_0^{-1}({\bf s},&{\bf k};P) \chi ({\bf k},{\bf k};P)= (2\pi)^3 \delta ({\bf s} - {\bf k}) + \tilde \chi ({\bf s},{\bf k};P) \Gamma ({\bf s}, {\bf k};P)
\end{eqnarray}
This can be solved using the Green's theorem and it gives
\begin{eqnarray}
   \chi ({\bf s},{\bf k};P)&= \chi_0 ({\bf s},{\bf k};P) + \int \frac{d{\bf q}}{(2\pi)^3} \, \chi_0 ({\bf s},{\bf q};P) \tilde \chi ({\bf q};P) \Gamma ({\bf q}, {\bf k};P).
\end{eqnarray}
Multiplying by $U_0({\bf s} - {\bf \tilde q})$ and integrating over ${\bf \tilde q}$
yields
\begin{eqnarray}
   \Gamma ({\bf s},{\bf k};P)= &\Gamma_0 ({\bf s},{\bf k};P) + \int \frac{d{\bf q}}{(2\pi)^3} \, \Gamma_0 ({\bf s},{\bf q};P) \tilde \chi ({\bf q};P) \Gamma ({\bf q}, {\bf k};P).
\label{eq:gamma}
\end{eqnarray}
This equation provides the mapping between $\Gamma$ and $\Gamma_0$. 

Assuming that the two-body scattering T-matrix $\Gamma_0({\bf s},{\bf k};P)$ does \emph{not} depend on
either of the two momenta ${\bf s}$ or ${\bf k}$ (but it may depend on the center-of-mass four-momentum $P$)
allows writing the many-body scattering T-matrix in the standard form
\begin{eqnarray}
   \Gamma (P)= \frac{\Gamma_0(P)}{1-\Gamma_0 (P) \int \frac{d{\bf q}}{(2\pi)^3} \, \tilde \chi ({\bf q};P)}.
\end{eqnarray}

\section*{References}

\bibliographystyle{unsrt}
\bibliography{ref}

\begin{thebibliography}{10}

\bibitem{Bloch2008a}
I.~Bloch, J.~Dalibard, and W.~Zwerger.
\newblock Many-body physics with ultracold gases.
\newblock {\em Rev. Mod. Phys.}, 80(3):885, 2008.

\bibitem{Holland2001a}
M.~Holland, S.~J. J. M.~F. Kokkelmans, M.~L. Chiofalo, and R.~Walser.
\newblock Resonance superfluidity in a quantum degenerate {F}ermi gas.
\newblock {\em Phys. Rev. Lett.}, 87(12):120406, 2001.

\bibitem{Ohashi2002a}
Y.~Ohashi and A.~Griffin.
\newblock {BCS-BEC} crossover in a gas of {F}ermi atoms with a {F}eshbach
  resonance.
\newblock {\em Phys. Rev. Lett.}, 89(13):130402, 2002.

\bibitem{Falco2005a}
G.~M. Falco and H.~T.~C. Stoof.
\newblock Atom-molecule theory of broad {F}eshbach resonances.
\newblock {\em Phys. Rev. A}, 71(6):063614, 2005.

\bibitem{Georges1996a}
A.~Georges, G.~Kotliar, W.~Krauth, and M.~J. Rozenberg.
\newblock Dynamical mean-field theory of strongly correlated fermion systems
  and the limit of infinite dimensions.
\newblock {\em Rev. Mod. Phys.}, 68(1):13, 1996.

\bibitem{Carlson2003a}
J.~Carlson, S.-Y. Chang, V.~R. Pandharipande, and K.~E. Schmidt.
\newblock Superfluid {F}ermi gases with large scattering length.
\newblock {\em Phys. Rev. Lett.}, 91(5):050401, 2003.

\bibitem{Vidal2003a}
G.~Vidal.
\newblock Efficient classical simulation of slightly entangled quantum
  computations.
\newblock {\em Phys. Rev. Lett.}, 91(14):147902, 2003.

\bibitem{Bulgac2009a}
A.~Bulgac and S.~Yoon.
\newblock Large amplitude dynamics of the pairing correlations in a unitary
  {F}ermi gas.
\newblock {\em Phys. Rev. Lett.}, 102(8):085302, 2009.

\bibitem{Chen2005a}
Q.~Chen, J.~Stajic, S.~Tan, and K.~Levin.
\newblock {BCS-BEC} crossover: From high temperature superconductors to
  ultracold superfluids.
\newblock {\em Phys. Rep.}, 412(1):1, 2005.

\bibitem{Giorgini2008a}
S.~Giorgini, L.~P. Pitaevskii, and S.~Stringari.
\newblock Theory of ultracold atomic {F}ermi gases.
\newblock {\em Rev. Mod. Phys.}, 80(4):1215, 2008.

\bibitem{Nozieres1985a}
P.~Nozi\`eres and S.~Schmitt-Rink.
\newblock {B}ose condensation in an attractive fermion gas: From weak to strong
  coupling superconductivity.
\newblock {\em J. Low Temp. Phys.}, 59(3):195, 1985.

\bibitem{Engelbrecht1992a}
J.~R. Engelbrecht and M.~Randeria.
\newblock Low-density repulsive {F}ermi gas in two dimensions: Bound-pair
  excitations and {F}ermi-liquid behavior.
\newblock {\em Phys. Rev. B}, 45(21):12419, 1992.

\bibitem{deMelo1993a}
C.~A.~R. S\'a~de Melo, M.~Randeria, and J.~R. Engelbrecht.
\newblock Crossover from {BCS} to {B}ose superconductivity: Transition
  temperature and time-dependent {G}inzburg-{L}andau theory.
\newblock {\em Phys. Rev. Lett.}, 71(19):3202, 1993.

\bibitem{Haussmann1994a}
R.~Haussmann.
\newblock Properties of a {F}ermi liquid at the superfluid transition in the
  crossover region between {BCS} superconductivity and {B}ose-{E}instein
  condensation.
\newblock {\em Phys. Rev. B}, 49(18):12975, 1994.

\bibitem{Perali2002a}
A.~Perali, P.~Pieri, G.~C. Strinati, and C.~Castellani.
\newblock Pseudogap and spectral function from superconducting fluctuations to
  the bosonic limit.
\newblock {\em Phys. Rev. B}, 66(2):024510, 2002.

\bibitem{Stajic2004a}
J.~Stajic, J.~N. Milstein, Q.~Chen, M.~L. Chiofalo, M.~J. Holland, and
  K.~Levin.
\newblock Nature of superfluidity in ultracold {F}ermi gases near {F}eshbach
  resonances.
\newblock {\em Phys. Rev. A}, 69(6):063610, 2004.

\bibitem{Haussmann2009a}
R.~Haussmann, M.~Punk, and W.~Zwerger.
\newblock Spectral functions and rf response of ultracold fermionic atoms.
\newblock {\em Phys. Rev. A}, 80(6):063612, 2009.

\bibitem{Tsuchiya2009a}
S.~Tsuchiya, R.~Watanabe, and Y.~Ohashi.
\newblock Single-particle properties and pseudogap effects in the {BCS-BEC}
  crossover regime of an ultracold {F}ermi gas above {$T_{c}$}.
\newblock {\em Phys. Rev. A}, 80(3):033613, 2009.

\bibitem{Hu2010a}
H.~Hu, X.-J. Liu, P.~D. Drummond, and H.~Dong.
\newblock Pseudogap pairing in ultracold {F}ermi atoms.
\newblock {\em Phys. Rev. Lett.}, 104(24):240407, 2010.

\bibitem{Watanabe2010a}
R.~Watanabe, S.~Tsuchiya, and Y.~Ohashi.
\newblock Superfluid density of states and pseudogap phenomenon in the
  {BCS-BEC} crossover regime of a superfluid {F}ermi gas.
\newblock {\em Phys. Rev. A}, 82(4):043630, 2010.

\bibitem{Gubbels2011a}
K.~B. Gubbels and H.~T.~C. Stoof.
\newblock Interacting preformed {C}ooper pairs in resonant {F}ermi gases.
\newblock {\em arXiv:1102.4751}, 2011.

\bibitem{Chiacchiera2009a}
S.~Chiacchiera, T.~Lepers, D.~Davesne, and M.~Urban.
\newblock Collective modes of trapped {F}ermi gases with in-medium interaction.
\newblock {\em Phys. Rev. A}, 79(3):033613, 2009.

\bibitem{Heiselberg2001a}
H.~Heiselberg.
\newblock {F}ermi systems with long scattering lengths.
\newblock {\em Phys. Rev. A}, 63(4):043606, 2001.

\bibitem{Magierski2009a}
P.~Magierski, G.~Wlaz\l{}owski, A.~Bulgac, and J.~E. Drut.
\newblock Finite-temperature pairing gap of a unitary {F}ermi gas by quantum
  {M}onte {C}arlo calculations.
\newblock {\em Phys. Rev. Lett.}, 103(21):210403, 2009.

\bibitem{Pekker2011a}
D.~Pekker, M.~Babadi, R.~Sensarma, N.~Zinner, L.~Pollet, M.~W. Zwierlein, and
  E.~Demler.
\newblock Competition between pairing and ferromagnetic instabilities in
  ultracold {F}ermi gases near {F}eshbach resonances.
\newblock {\em Phys. Rev. Lett.}, 106(5):050402, 2011.

\bibitem{Diener2008a}
R.~B. Diener, R.~Sensarma, and M.~Randeria.
\newblock Quantum fluctuations in the superfluid state of the {BCS-BEC}
  crossover.
\newblock {\em Phys. Rev. A}, 77(2):023626, 2008.

\bibitem{Eagles1969}
D.~M. Eagles.
\newblock Possible pairing without superconductivity at low carrier
  concentrations in bulk and thin-film superconducting semiconductors.
\newblock {\em Phys. Rev.}, 186(2):456, 1969.

\bibitem{Leggett}
A.~J. Leggett.
\newblock In {\em Modern Trends in the Theory of Condensed Matter}. Editors
  Pekalski, A. and Przystawa, R. Springer-Verlag, Berlin, 1980.

\bibitem{Papp2008a}
S.~B. Papp, J.~M. Pino, R.~J. Wild, S.~Ronen, C.~E. Wieman, D.~S. Jin, and
  E.~A. Cornell.
\newblock {B}ragg spectroscopy of a strongly interacting {$^{85}Rb$}
  {B}ose-{E}instein condensate.
\newblock {\em Phys. Rev. Lett.}, 101(13):135301, 2008.

\bibitem{Kinnunen2009a}
J.~J. Kinnunen and M.~J. Holland.
\newblock {B}ragg spectroscopy of a strongly interacting {B}ose-{E}instein
  condensate.
\newblock {\em New J. Phys.}, 11:013030, 2009.

\bibitem{Ronen2009a}
S.~Ronen.
\newblock The dispersion relation of a {B}ose gas in the intermediate- and
  high-momentum regimes.
\newblock {\em J. Phys. B:At. Mol. Opt. Phys.}, 42:055301, 2009.

\bibitem{Cazalilla2010a}
M.~A. Cazalilla.
\newblock A composite fermion approach to the ultracold dilute {F}ermi gas.
\newblock {\em arxiv:1005.1363}, 2010.

\bibitem{FetterAndWalecka}
A.~L. Fetter and J.~D. Walecka.
\newblock {\em Quantum theory of many-particle systems}.
\newblock McGraw-Hill, New York, 1971.

\bibitem{Luttinger1961a}
J.~M. Luttinger.
\newblock Analytic properties of single-particle propagators for many-fermion
  systems.
\newblock {\em Phys. Rev.}, 121(4):942, 1961.

\bibitem{Lobo2006a}
C.~Lobo, A.~Recati, S.~Giorgini, and S.~Stringari.
\newblock Normal state of a polarized {F}ermi gas at unitarity.
\newblock {\em Phys. Rev. Lett.}, 97(20):200403, 2006.

\bibitem{Chevy2006a}
F.~Chevy.
\newblock Universal phase diagram of a strongly interacting {F}ermi gas with
  unbalanced spin populations.
\newblock {\em Phys. Rev. A}, 74(6):063628, 2006.

\bibitem{Combescot2007a}
R.~Combescot, A.~Recati, C.~Lobo, and F.~Chevy.
\newblock Normal state of highly polarized {F}ermi gases: Simple many-body
  approaches.
\newblock {\em Phys. Rev. Lett.}, 98(18):180402, 2007.

\bibitem{Prokofev2008a}
N.~Prokof'ev and B.~Svistunov.
\newblock {F}ermi-polaron problem: Diagrammatic {M}onte {C}arlo method for
  divergent sign-alternating series.
\newblock {\em Phys. Rev. B}, 77(2):020408, 2008.

\bibitem{Combescot2008a}
R.~Combescot and S.~Giraud.
\newblock Normal state of highly polarized {F}ermi gases: Full many-body
  treatment.
\newblock {\em Phys. Rev. Lett.}, 101(5):050404, 2008.

\bibitem{Punk2009a}
M.~Punk, P.~T. Dumitrescu, and W.~Zwerger.
\newblock Polaron-to-molecule transition in a strongly imbalanced {F}ermi gas.
\newblock {\em Phys. Rev. A}, 80(5):053605, 2009.

\bibitem{Shin2008a}
Y.-i. Shin.
\newblock Determination of the equation of state of a polarized {F}ermi gas at
  unitarity.
\newblock {\em Phys. Rev. A}, 77(4):041603, 2008.

\bibitem{Schirotzek2009a}
A.~Schirotzek, C.-H. Wu, A.~Sommer, and M.~W. Zwierlein.
\newblock Observation of {F}ermi polarons in a tunable {F}ermi liquid of
  ultracold atoms.
\newblock {\em Phys. Rev. Lett.}, 102(23):230402, 2009.

\bibitem{Nascimbene2009a}
S.~Nascimb\`ene, N.~Navon, K.~J. Jiang, L.~Tarruell, M.~Teichmann, J.~McKeever,
  F.~Chevy, and C.~Salomon.
\newblock Collective oscillations of an imbalanced {F}ermi gas: Axial
  compression modes and polaron effective mass.
\newblock {\em Phys. Rev. Lett.}, 103(17):170402, 2009.

\bibitem{Nascimbene2010a}
S.~Nascimb\`ene, N.~Navon, K.~J. Jiang, F.~Chevy, and C.~Salomon.
\newblock Exploring the thermodynamics of a universal {F}ermi gas.
\newblock {\em Nature}, 463:1057, 2010.

\bibitem{Pilati2008a}
S.~Pilati and S.~Giorgini.
\newblock Phase separation in a polarized {F}ermi gas at zero temperature.
\newblock {\em Phys. Rev. Lett.}, 100(3):030401, 2008.

\bibitem{Combescot2009a}
R.~Combescot, F.~Alzetto, and X.~Leyronas.
\newblock Particle distribution tail and related energy formula.
\newblock {\em Phys. Rev. A}, 79(5):053640, 2009.

\bibitem{Tan2008a}
S.~Tan.
\newblock Energetics of a strongly correlated {F}ermi gas.
\newblock {\em Ann. Phys.}, 323:2952, 2008.

\bibitem{Tan2008b}
S.~Tan.
\newblock Large momentum part of a strongly correlated {F}ermi gas.
\newblock {\em Ann. Phys.}, 323:2971, 2008.

\bibitem{Sartor1980a}
R.~Sartor and C.~Mahaux.
\newblock Self-energy, momentum distribution, and effective masses of a dilute
  {F}ermi gas.
\newblock {\em Phys. Rev. C}, 21(4):1546, 1980.

\bibitem{Gandolfi2010a}
S.~Gandolfi, K.~E. Schmidt, and J.~Carlson.
\newblock {BEC-BCS} crossover and universal relations in unitary {F}ermi gases.
\newblock {\em arxiv:1012.4417}, 2010.

\bibitem{Navon2010a}
N.~Navon, S.~Nascimbène, F.~Chevy, and C.~Salomon.
\newblock The equation of state of a low-temperature {F}ermi gas with tunable
  interactions.
\newblock {\em Science}, 328(5979):729, 2010.

\bibitem{Kuhnle2010a}
E.~D. Kuhnle, H.~Hu, X.-J. Liu, P.~Dyke, M.~Mark, P.~D. Drummond, P.~Hannaford,
  and C.~J. Vale.
\newblock Universal behavior of pair correlations in a strongly interacting
  {F}ermi gas.
\newblock {\em Phys. Rev. Lett.}, 105(7):070402, 2010.

\bibitem{Gaebler2010a}
J.~P. Gaebler, J.~T. Stewart, T.~E. Drake, D.~S. Jin, A.~Perali, P.~Pieri, and
  G.~C. Strinati.
\newblock Observation of pseudogap behaviour in a strongly interacting {F}ermi
  gas.
\newblock {\em Nat. Phys.}, 6:569, 2010.

\bibitem{Leskinen2010a}
M.~J. Leskinen, J.~Kajala, and J.~J. Kinnunen.
\newblock Resonant scattering effect in spectroscopies of interacting atomic
  gases.
\newblock {\em New J. Phys.}, 12:083041, 2010.

\bibitem{Ohashi2005a}
Y.~Ohashi and A.~Griffin.
\newblock Single-particle excitations in a trapped gas of {F}ermi atoms in the
  {BCS-BEC} crossover region.
\newblock {\em Phys. Rev. A}, 72(1):013601, 2005.

\bibitem{Jensen2007a}
L.~M. Jensen, J.~Kinnunen, and P.~T\"orm\"a.
\newblock Non-{BCS} superfluidity in trapped ultracold {F}ermi gases.
\newblock {\em Phys. Rev. A}, 76(3):033620, 2007.

\end{thebibliography}

\end{document}